\documentclass[sigconf, nonacm]{acmart}

\makeatletter
\def\@ACM@checkaffil{% Only warnings
    \if@ACM@instpresent\else
    \ClassWarningNoLine{\@classname}{No institution present for an affiliation}%
    \fi
    \if@ACM@countrypresent\else
        \ClassWarningNoLine{\@classname}{No country present for an affiliation}%
    \fi
}
\makeatother

\AtBeginDocument{%
  \providecommand\BibTeX{{%
    \normalfont B\kern-0.5em{\scshape i\kern-0.25em b}\kern-0.8em\TeX}}}

\usepackage{comment}
\usepackage{todonotes}
\usepackage{xspace}
\usepackage[flushleft]{threeparttable}
\usepackage{enumitem}
\usepackage{balance}
\usepackage{graphics}
\usepackage{color, colortbl, soul}
\usepackage{tikz}
\usepackage{caption,subcaption}
\usepackage{fixltx2e}
\usepackage{bm}
\usepackage{romannum}
\usepackage{mathtools}
\usepackage{tabularx, booktabs, caption, ragged2e}
\usepackage{tikzsymbols}
\usepackage{amsmath}
\usepackage[title]{appendix}
\usepackage{bm}
\usepackage{algorithm}
\usepackage{algorithmic}
\usepackage[normalem]{ulem}

\newcommand\algorithmicprocedure{\textbf{function}}
\newcommand{\algorithmicendprocedure}{\algorithmicend\ \algorithmicprocedure}
\makeatletter
\newcommand\PROCEDURE[3][default]{%
  \ALC@it
  \algorithmicprocedure\ \textsc{#2}(#3)%
  \ALC@com{#1}%
  \begin{ALC@prc}%
}
\newcommand\ENDPROCEDURE{%
  \end{ALC@prc}%
  \ifthenelse{\boolean{ALC@noend}}{}{%
    \ALC@it\algorithmicendprocedure
  }%
}
\newenvironment{ALC@prc}{\begin{ALC@g}}{\end{ALC@g}}
\makeatother

\def\S{\mbox{Section}\xspace}
\def\A{\mbox{Appendix}\xspace}

\def\alt{{\mbox{alt}}\xspace}

\def\ie{\textit{i.e.,}\xspace}
\def\etal{\textit{et~al.}\xspace}

\def\eg{\textit{e.g.,}\xspace}
\def\incl{\textit{incl.}\xspace}

\def\vs{\textit{vs.}\xspace}

\def\First{\textit{First}\xspace}
\def\Second{\textit{Second}\xspace}

\definecolor{redcolor}{RGB}{255,0,0}

\definecolor{olivegreen}{RGB}{107,142,35}
\definecolor{brickred}{RGB}{203,65,84}
\definecolor{darkpink}{RGB}{76, 0, 153}
\definecolor{lightblue}{RGB}{0, 128, 255}

\colorlet{transparentred}{red!20}
\colorlet{transparentgreen}{green!20}
\DeclareRobustCommand{\hlred}[1]{{\sethlcolor{transparentred}\hl{#1}}}
\DeclareRobustCommand{\hlgreen}[1]{{\sethlcolor{transparentgreen}\hl{#1}}}

\newcommand\lword[1]{\leavevmode\nobreak\hskip0pt plus\linewidth\penalty50\hskip0pt plus-\linewidth\nobreak #1}

\newcommand\inlinecode[2][]{\lword{\tikz[overlay]\node[fill=blue!10, inner sep=1pt, anchor=text, rectangle, rounded corners=1mm,#1] {\ttfamily #2};\phantom{\ttfamily #2}}}

\DeclareMathOperator*{\argmax}{argmax}

\setcopyright{none}

\author{Chen Chen}
\orcid{0000-0001-7179-0861}
\affiliation{%
  \institution{Florida International University}
  \city{Miami}
  \state{FL}
  \country{United States}
}
\email{chechen@fiu.edu}

\author{Cuong Nguyen}
\orcid{0000-0001-9234-9960}
\affiliation{%
  \institution{Adobe Research}
  \city{San Francisco}
  \state{CA}
  \country{United States}
}
\email{cunguyen@adobe.com}

\author{Alexa Siu}
\orcid{0000-0002-4879-1476}
\affiliation{%
  \institution{Adobe Research}
  \city{San Jose}
  \state{CA}
  \country{United States}
}
\email{asiu@adobe.com}

\author{Dingzeyu Li}
\orcid{0000-0002-4222-8105}
\affiliation{%
  \institution{Adobe Research}
  \city{Seattle}
  \state{WA}
  \country{United States}
}
\email{dinli@adobe.com}

\author{Nadir Weibel}
\orcid{0000-0002-3457-4227}
\affiliation{%
  \institution{University of California San Diego}
  \city{La Jolla}
  \state{CA}
  \country{United States}
}
\email{weibel@ucsd.edu}

\renewcommand{\shortauthors}{Chen~\etal}


\begin{document}

\def\sysname{SweeperBot}

%TC:ignore

\title{\sysname: Making 3D Browsing Accessible through View Analysis and Visual Question Answering}

\begin{abstract}
Accessing 3D models remains challenging for \textbf{S}creen \textbf{R}eader~(SR) users.
While some existing 3D viewers allow creators to provide alternative text, they often lack sufficient detail about the 3D models.
Grounded on a formative study, this paper introduces \emph{\sysname}, a system that enables SR users to leverage visual question answering to explore and compare 3D models.
\sysname~answers SR users' visual questions by combining an optimal view selection technique with the strength of generative- and recognition-based foundation models.
An expert review with $10$~\textbf{B}lind and \textbf{L}ow-\textbf{V}ision~(BLV) users with SR experience demonstrated the feasibility of using \sysname~to assist BLV users in exploring and comparing 3D models.
The quality of the descriptions generated by \sysname~was validated by a second survey study with $30$~sighted participants.

\end{abstract}

\pagenumbering{arabic}
\settopmatter{printfolios=true}
\maketitle

%TC:endignore
\begin{figure*}
    \centering
    \includegraphics[width=\textwidth]{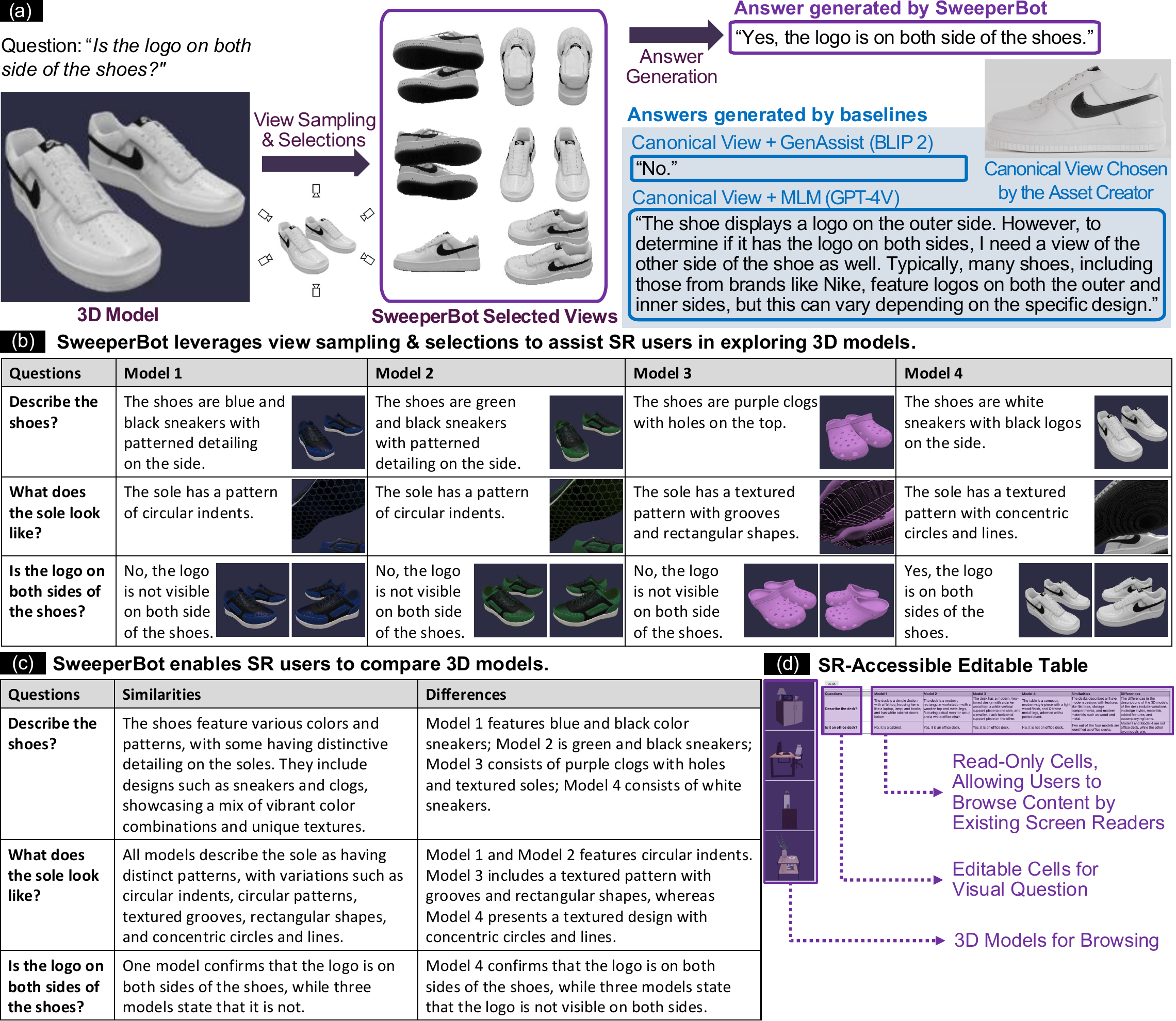}
    \caption{(a) \sysname~ generates descriptions based on \textbf{B}lind and \textbf{L}ow \textbf{V}ision (BLV) users' visual questions (top left). The description created by \sysname ~ ({\color{darkpink}pink box}, right) more accurately answers the visual question, compared to the baselines ({\color{lightblue}blue boxes}, right) using the canonical view chosen by the creator. (b - c)~The generated table supports BLV users to navigate the generated descriptions with existing \textbf{S}creen \textbf{R}eaders~(SRs). (d) \sysname's interface with an SR-accessible editable table.}
    \label{fig::accessible_table}
\end{figure*}

\section{Introduction}\label{sec::intro}

3D models are crucial in many real-world applications, valued for their intuitiveness and interactivity. 
When browsing 3D content online, users often need to view and compare multiple 3D models. 
For example, while purchasing furniture from IKEA, customers can use the ``view in 3D'' feature to examine and compare 3D models of products directly in the browser~\cite{Ikea3DViewer}.
However, this process heavily relies on \emph{visual} perception, making it challenging for \textbf{B}lind and \textbf{L}ow-\textbf{V}ision (BLV) users.

Technologies like \textbf{S}creen \textbf{R}eaders (SRs) are widely used by BLV users to access non-textual content by vocalizing the \textbf{alt}ernative (\alt) texts~\cite{alttext}.
Some 3D viewers (\eg~\cite{ModelViewerGoogle, Babylonjs}) support the inclusion of \alt text, providing a description of the rendered 3D model.
However, this is insufficient because the \alt~texts may not include the key information that BLV users need~\cite{Winkle2020}. 
For example, sellers of the sneakers shown in Figure~\ref{fig::accessible_table}a may only provide descriptions of the overall design style while omitting details like the design patterns and the bottom sole styles.
Yet, this detailed information is essential for BLV customers to fully understand the aesthetics and functionality of the shoes, enabling them to make informed purchasing decisions.
Few 3D viewers like Babylon.js~\cite{BabylonAccessibility2023} allow creators to write \alt texts for selected key objects, which SRs can vocalize based on cursor interactions.
But creating them is time-consuming and tedious.  
Consuming these \alt texts also requires BLV users to navigate views using a mouse, which is frustrating and inefficient, as keyboard shortcuts are often considered as primary input and browsing strategies for navigating focused elements with mainstream SRs~\cite{Borodin2010}.

Automatic \textbf{V}isual \textbf{Q}uestion \textbf{A}nswering (VQA) \cite{Antol2015} systems like GenAssist \cite{Huh2023} are promising to aid BLV users in accessing visual content. 
These interfaces that are often powered by \textbf{V}ision \textbf{L}anguage \textbf{F}oundation \textbf{M}odels (VLFMs) allow users to ask questions about an image. 
It is not trivial to extend current VQA systems to support 3D understanding. 
These systems typically require an image as input, meaning a rendered camera view of the 3D model must be selected beforehand.
This ``\textit{canonical view}'' approach is undesirable, as a 3D model can be viewed from countless perspectives. 
There might be important details at the back or at the bottom that are not covered by the canonical view. 
Without the context of additional views, VLFMs would lack sufficient information to generate useful responses to the user’s question about the 3D model, leading to hallucinations (Figure~\ref{fig::accessible_table}a).
It remains challenging to extend VQA systems to support the multi-view nature of 3D models.

To understand the current practices and users' expectations regarding accessing 3D models, a formative study was conducted by interviewing two blind users with 3D and SR experience.
With three key considerations, \emph{\sysname}~was prototyped as a browser-based application with an editable table (Figure~\ref{fig::accessible_table}d).
BLV users can compose free-form visual questions or commands to query about the overall content and/or details of 3D models.
\sysname~then analyzes the visual questions and the 3D model, followed by updating the answers in the editable table.
This is achieved through a novel VQA pipeline, including \emph{view sampling}, \emph{view selections}, and \emph{answer generations}.
Figure~\ref{fig::accessible_table}a illustrates how \sysname~assists BLV users in accessing 3D models retrieved from the ``shoes'' catalog. 
Unlike descriptions generated using the canonical view, \sysname~can accurately understand the logo design of the shoes by leveraging the selected relevant views.
BLV users can use existing SRs to hear, navigate, and access the table. 
With generated descriptions, BLV users can ask follow-up questions to gain a full mental understanding of the 3D models (Figure~\ref{fig::accessible_table}b - c).

An expert review with $10$~BLV users with SR experience demonstrated the feasibility and effectiveness of using \sysname~to assist BLV users in exploring and comparing 3D models.
A second survey study with $30$~sighted users validated the quality of the \sysname-generated descriptions compared to baselines relying on the canonical views selected by the creators of 3D assets.

\section{Related Work}\label{sec::related}

\subsection{Accessing 3D Models and Scenes via Audio}\label{sec::related::3d}

Using audible speech is promising to help BLV users access 3D models and scenes without requiring additional hardware.
Even for low-vision users, text-to-speech and auditory systems remain widely used despite their residual vision~\cite{Szpiro2016, Crossland2014}.

\textbf{S}creen \textbf{R}eaders (SRs) are widely used by people who are blind, visually impaired, or may otherwise struggle to read on-screen content~\cite{Edwards2023, ScreenReaders, Szpiro2016}.
Winkle~\cite{Winkle2020} speculated on a set of usability expectations that SRs need to consider when trying to make 3D content accessible.
However, requiring BLV users to rely on the mouse to navigate the camera as sighted users do is challenging and frustrating~\cite{Borodin2010}.
A few 3D viewers (\eg~\cite{ModelViewerGoogle, BabylonAccessibility2023}) let developers to specify \emph{what}, \emph{when}, and \emph{how} the \alt texts should be announced, but they often lack crucial details that BLV users need.

Recent scene description smartphone applications allow BLV users to ask visual questions related to surroundings - a real-world 3D content equivalent captured by rear-facing camera.
VizWiz introduced a mobile application that enables BLV users to upload images and visual questions, which are then answered by the crowd~\cite{Bigham2010VizWiz}.
Visual interpretation services like \mbox{Aira}~\cite{aria} and \mbox{CrowdViz}~\cite{Crowdviz2015} provide real-time video-audio connections between BLV users and sighted assistants.
Similarly, commercial application Be My Eyes \cite{bemyeyes} has been used by $300$K+ BLV users and $4.7$M volunteers.
Researchers also explored integrating automatic VQA pipelines, moving beyond sole reliance on the crowd, a topic that will be discussed in \S~\ref{sec::related::aialt}.
Similar to these scene description applications, \sysname~uses an automatic VQA approach with existing SRs, enabling BLV users to type visual questions and browse AI-generated descriptions to explore and compare 3D models.

\subsection{Automatic VQA for 3D Models and Scenes}\label{sec::related::vqa}
VQA expands the traditional text-based question answering~\cite{Rajpurkar2016} by integrating visual content, allowing for image(s) interpretations~\cite{Antol2015}.
\textbf{M}ultimodal \textbf{L}arge \textbf{L}anguage \textbf{M}odels~(MLLMs) like \mbox{GPT-4} \cite{openai2024gpt4, GPTVision, GPT4o, GPT45} and \mbox{GPT-5} \cite{GPT5} allow multiple images to be used as prompts alongside text.
\sysname~extends these ideas to 3D and shows how to generate descriptions from selected sampled views.
Extending existing image-based VQA pipelines to help BLV users access 3D models is challenging:
Identifying relevant views for visual questions is difficult for BLV users due to limited visual abilities;
although canonical view(s) chosen by 3D model creators may be helpful, some visual questions may be difficult to answer using these views.

VLFMs have been used to caption 3D models.
Cap3D~\cite{Luo2023} showed a captioning pipeline by combining BLIP-generated captions for uniformly sampled views with GPT-4.
While not explicitly described, a similar approach could be used for VQA by prompting MLLMs like GPT-4V~\cite{openai2024gpt4} with all sampled views.
DiffuRank \cite{Luo2024ViewSelection} demonstrated how sampled views can be ranked based on the alignment between their captions and the corresponding 3D models, aiming to minimize Cap3D’s hallucinations by selecting views that can better represent the key characteristics of the 3D model.
ShapeLLM~\cite{Qi2024ShapeLLM} designed a pipeline to understand point clouds by leveraging features from multi-view images.
VFC~\cite{Ge2024VFC} used a dedicated object-detection tool for LLM to fact-check view-dependent captions.
If important objects identified by the LLM are missed, VFC will discard the corresponding captions for subsequent final caption synthesis.
Unlike captioning, VQA requires AI to focus on multiple optimal and relevant views regarding visual questions.
Generating answers using less good and irrelevant views can cause challenges such as missing details and content hallucinations.

Researchers also explored VQA pipelines for 3D environments.
Most works~\cite{Ye2024, Etesam2023, Azuma2022ScanQA, Linghu2024, Yin2023LAMM, Yang2024} rely on the annotated dataset that aligns 3D point clouds and textual queries.
Singh~\etal \cite{Singh2024} evaluated the performance of GPT-based agents on 3D VQA benchmarks, focusing on the spatial relations between different objects.
It is unclear how well this approach can be extended to enhance the accessibility of 3D models for BLV users by addressing their visual inquiries, which may encompass descriptions of shapes, materials, and design styles.

\sysname~introduces a zero-shot VQA pipeline, including \emph{view sampling}, \emph{selection}, and \emph{answer generations}.
While BridgeQA~\cite{Mo2024BridgeQA} demonstrated how to accomplish 3D VQA tasks using the question-conditioned 2D view selected from video frames rendering the 3D scene, solely relying on BLIP's image-text retrieval model~\cite{Li2022blip} is insufficient, as views of the 3D model may differ from the 2D images used to train the BLIP model.
Therefore, evaluating view significance using cues from key objects is essential.
Additionally, many visual questions, \eg~describing the logo design of the shoes in Figure~\ref{fig::accessible_table}b - c, are challenging to address using single 2D view.
Instead of emphasizing captioning like~\cite{Luo2023, Ge2024VFC}, \sysname~focuses on VQA by analyzing the selected relevant views of 3D models related to the visual questions.
BLV users are enabled to access finer details of 3D models by sampling the distance between the viewing camera and the 3D model as well as assessing the importance of the sampled views based on key objects.

\subsection{Enhancing Visual Accessibility of Images and 3D using Automatic VQA Pipelines}\label{sec::related::aialt}
Researchers have explored how to integrate automatic VQA pipelines to help BLV users access images and 3D.
For example, Revamp~\cite{Wang2021} proposed an interface that allows SR users to ask questions while browsing an e-commerce website.
ImageAssist~\cite{Vishnu2023} integrated an automatic VQA pipeline to enhance visual accessibility for images rendered on touchscreens.
GenAssist~\cite{Huh2023} presented an interface that enables BLV users to access the searched and AI-generated images using BLIP~\cite{Li2022blip} and GPT-4~\cite{openai2024gpt4}.
VizAbility~\cite{Gorniak2024} demonstrated an LLM-enabled pipeline where SR users could query visual, analytical, contextual, and navigational questions for charts.

When it comes to 3D, AI-powered scene description applications like Seeing AI~\cite{seeingAi} allow BLV users to ask questions related to the views captured by mobile cameras using multiple task-specific vision models. 
VizWiz~\cite{Gurari2018, Gurari2019, VizWizDataset} demonstrated a dataset created entirely by BLV individuals that can be used to develop an automated VQA pipeline.
Such automated image-based VQA features have also been integrated into extended reality systems like SeeingVR \cite{Zhao2019SeeingVR}, allowing low-vision users to better access specific views within virtual environments.
However, requiring BLV users to identify relevant viewpoints is challenging. 
%
% why seeing VR's approach cannot be used for our cases.
While low-vision users may leverage residual vision to select relevant views in SeeingVR~\cite{Zhao2019SeeingVR}, this approach becomes challenging for users without usable vision.
Less optimal views can cause misleading AI inference results~\cite{Bigham2011}.
Similar findings grounded on Seeing AI~\cite{seeingAi} have also been unveiled in our formative study (\S~\ref{sec::formative::results}).
Although recent works like VizWiz::LocateIt ~\cite{Bigham2010} and EasySnap~\cite{White2010} proposed novel approaches to interactively guide BLV users to locate objects and find high-quality viewpoints using computer vision techniques and auditory feedback, integrating such techniques into \sysname~is challenging:
Requiring BLV users to find relevant views is impractical and time-consuming;
unlike images, many visual questions related to 3D models can only be addressed by examining and synthesizing multiple views.
VRSight~\cite{Killough2025VRSight} demonstrated how a set of task-specific AI models can be integrated to help blind users explore key elements (e.g., tables, avatars) in virtual reality; however, accessing fine-grained details remains challenging.

\section{Formative Study}\label{sec::formative}
Semi-structured interviews \cite{Adams2015} were conducted with BLV users with SR and 3D experience to understand the current practices for how BLV users access 3D models, the challenges they face, and their expectations for an AI tool to help access 3D models.

\subsection{Participants and Procedures}\label{sec::formative::participant}
Participants with prior SR and 3D experience are intended to be recruited.
However, this is challenging, as accessing 3D models solely based on existing SRs is impractical.
Therefore, we recruited participants with 3D printing experience, which often requires BLV users to explore and compare 3D models by touching tangible 3D-printed artifacts.
Two blind users were recruited as the \textbf{F}ormative study \textbf{P}articipants (FP\#, age $M = 28.5$ $SD = 4.95$, see Appendix~\ref{sec::app::blv}).
FP1 and FP2 have three and six months of 3D printing experience, respectively, with prior experience exploring online 3D models.
Participants considered themselves experienced users of SRs.
Two guiding questions were used in the interview: {\it ``what does the experience of exploring and comparing 3D models look like?''} and {\it ``what are the limitations of using SRs to access 3D models?''}
Thematic analysis~\cite{Braun2012} and a mixture of emergent and priori coding~\cite{Lazar2017} were used to analyze the qualitative data.
All studies have been approved by the \textbf{I}nstitutional \textbf{R}eview \textbf{B}oard~(IRB)~(see \A~\ref{sec::app::irb} for more details).

\subsection{Results}\label{sec::formative::results}

\noindent{\bf How do BLV users explore and compare 3D models? }

\noindent Participants reported exploring and comparing 3D content with existing tools is challenging: {\it ``[While finding 3D models for printing] it's very hard to view objects and things as a blind person''} (FP1), and {\it ``3D model viewers are still only for people who are sighted''}~(FP2).
These testimonies are similar to Siu~\etal's findings~\cite{Siu2019ShapeCAD} while integrating tactile display into the 3D modeling workflow.
Participants indicate two strategies to access 3D models:

\vspace{+4px}\noindent $\bullet$~{\bf Profile texts.}
When selecting a 3D model using textual queries, participants primarily depend on the profile texts, which could be vocalized by SRs, and then proceed to 3D print the model to confirm their expected extrapolations.
For example, FP1 described:
{\it ``What I usually do is to go to a website like Thingiverse\footnote{Thingiverse: \url{https://www.thingiverse.com}. Accessed on January 10, 2025.} that has profiles for each model. I would begin by reviewing the profile texts for each of the models recommended by the website. Then I usually just send them to my printer and print them, and hope that they turn out to be what they're supposed to [...] There's not a good way to visualize what you're printing before you print it.''}

\vspace{+4px}\noindent $\bullet$~{\bf LLM and sighted friends.}
Participants also used the captioning capabilities of recent LLMs to access 3D models.
FP2 explained: {\it ``What I do right now is a combination of using ChatGPT and using my sighted friends. So, what I'll do is search for some 3D models. Then, I will take screenshots of the rendered view. And then I'll put it into ChatGPT to have a description of what is going on. But most of the time, it's still not giving me the details that I got from when I printed them out with my 3D printer. So I have to check with my sighted friends.''}~
While both approaches help BLV users consume and compare 3D models, participants acknowledged the workflows to be laborious, tedious, and often less reliable.
While FP2 might ask for help from sighted users, FP1 held an opposite opinion and emphasized the importance of \emph{independence}. She emphasized: {\it ``Family and friends are not always around [...] We're not always going to have our people with us, who can help us to read things.''}

\vspace{+4px}\noindent{\bf Navigating and identifying relevant views is challenging.}
While accessing 3D models for sighted users requires \emph{active} interaction, where users need to navigate the viewing camera and explore different perspectives~\cite{Winkle2020}, participants agreed that such active interactions are difficult for BLV users.
Although not having experience of navigating the viewing camera for consuming 3D models, FP1 noted her experience of using Seeing AI~\cite{seeingAi}: 
{\it ``You have to take a picture of what you want described [...] But for me, taking pictures is not fun or easy, because I simply could not see the camera view as sighted people.''}
FP2 faced similar obstacles while employing ChatGPT to interpret the captured screenshot of the 3D model, and the screenshot failed to provide a clear perspective.
Having experience of using AstroPrint\footnote{AstroPrint: \url{https://www.astroprint.com}. Accessed on January 10, 2025.}, FP2 suggested the helpfulness to have the SRs vocalize critical information of 3D models without forcing BLV users to interact with the viewing camera: {\it ``It's really hard to know what the model will look like in different positions and orientations. Also, I don't get any confirmation after changing the view. Ideally, screen readers should just read all the necessary information that I need without needing to change different view angles.''}
FP2 also emphasized the inaccessibility of the 3D viewer in AstroPrint: {\it ``[Despite the accessible features of various menus] all of this comes down to the point that I can see the model by navigate and explore different views.''}

\vspace{+4px}\noindent{\bf Support of accessing requested details.}
Participants emphasized the importance of accessing the requested details, \eg~{\it ``What would be revolutionary is if there was an app where you could just import whatever models you're wanting. Let the app analyze it. And then ask your questions''}~(FP1). 
While FP2 found ChatGPT to be useful, he emphasized the need for accessing more specific details, as {\it``[the descriptions from the ChatGPT] still not giving me the details that I got from when I printed it out''}.
For example, {\it ``I'm working on an iPhone case. But sometimes ChatGPT will say something like, `Okay, the case appears to have cutouts for the camera and the speakers and all that'. And then I will ask the AI: `Can you see if the cutouts go all the way through?', `Does it create a hole?'''} (FP2).
Despite not having AI experience, FP1 emphasized the importance of having accurate AI answers: {\it ``It needs to make sure that its accuracy is like 99\% to 100\% accurate as far as giving details. [FP1 then used the 3D viewer example for online shopping experience] We're not buying the wrong thing, because the app describes the wrong thing.''}
Although using profile texts helps, FP1 still complained: {\it ``When I 3D print something, I don't know what it's going to turn out like until it's done.''}~
While using a ChatGPT-like AI method as FP2 might be useful, FP1 believes {\it ``it is important to make sure that its accuracy is high enough for the requested details''}.

\subsection{Design Considerations}\label{sec::considerations}
The formative study demonstrated the current practices, challenges, and expectations from BLV users for accessing 3D content, which has informed the design of \sysname.
Despite the limited sample size, the formative nature of the study, and the fact that participants’ 3D experience was primarily in 3D printing, many of their insights and prior experiences can be generalized to broader contexts such as online shopping as emphasized by FP1.
Three key \textbf{D}esign \textbf{C}onsiderations (DC) are summarized:

\vspace{+4px}\noindent{\bf (DC1) Assistance for exploring and comparing 3D models.}
Our findings suggested that it is crucial to both \emph{explore} individual 3D models and \emph{compare} them.
This shares similar findings with GenAssist~\cite{Huh2023}, focusing on AI-generated images. BLV users' expectation for 3D model exploration aligns well with their 2D expectations.

\vspace{+4px}\noindent{\bf (DC2) Providing reliable responses to questions about specific details.}
The principles of \emph{overview} and \emph{details} are not solely relevant to how sighted users seek visual information~\cite{Shneiderman1996, Cockburn2009}, they also apply to BLV users when accessing 3D models.
While participants described their current strategies to access and explore 3D models, we revealed the inefficiencies due to the lack of critical details, such as the specific shapes, color, and textures.

\vspace{+4px}\noindent{\bf (DC3) Encapsulating the complexities of navigating and exploring views. }
While sighted users typically need to \emph{actively} interact to access 3D content~\cite{Winkle2020}, we found that creating \emph{passive} methods for BLV users to access 3D content is essential.
It is challenging for BLV users to navigate and explore different views.
Therefore, the design of \sysname~should prevent view-dependent \alt~text that might require BLV users to \emph{actively} navigate the viewing camera to access the details of 3D models.
\section{\sysname}\label{sec::system}
\sysname~aims to facilitate BLV users to explore and compare 3D models.
As a first step, \sysname~\emph{only} focuses on \emph{simple} 3D models that only need to be examined \emph{externally}, rather than complex and large 3D environments.
The desk model (Figure~\ref{fig::study_tasks}b) will be used as the running example.
\A~\ref{sec::app::implementations} details implementations.

\subsection{Editable and SR-Accessible Table}\label{sec::sys::table}
While navigating the viewing camera is the primary method for sighted users to access 3D models, \textbf{DC3} highlights the difficulties that BLV users face when trying to interact with 3D views.
\sysname~was designed as an editable and SR-accessible table (Figure ~\ref{fig::accessible_table}d). 
SR users can type the questions to query the overall content or specific details of 3D models without interacting with the viewing cameras.
\sysname's VQA pipeline then generates responses for individual 3D models while also summarizing their similarities and differences.

\sysname~allows BLV users to access \emph{four} preloaded 3D models given the visual questions.
This design was based on the approximate visual memory capacity for sighted users~\cite{Owen2004, Vogel2004}.
Similar ideas of generating four images simultaneously have been adopted in existing \textbf{G}enerative \textbf{AI}~(GenAI) tools, \eg~\cite{Firefly, Huh2023}.
\sysname~focuses on accessing individual 3D models and making comparisons (\textbf{DC1}).
The spirit of textualizing visual content into an SR-accessible table using the interactive VQA is similar to GenAssist~\cite{Huh2023}, which focuses on image accessibility. 
BLV users can then navigate the table cells using the browse mode of existing SRs\footnote{Mainstream PC SRs offer \emph{browse} and \emph{focus} modes to interact with GUI applications. \emph{Browse} mode enables SRs to navigate every GUI element, including those that are not focusable by keyboard. \emph{Focus} mode allows users to interact with the focusable GUI element like the text field~\cite{BrowseFocusMode}.}.

\subsection{VQA Pipeline}\label{sec::sys::vqa}

\sysname~needs to support inferring reliable responses to the questions related to the overview or the specific details of the 3D models (\textbf{DC2}).
A novel VQA pipeline was designed that brings the strengths of generative and task-specific recognition-based foundation models to infer answers to visual questions posed by BLV users.
\sysname's pipeline includes three stages. 

\begin{figure*}[t]
    \centering
    \includegraphics[width=\textwidth]{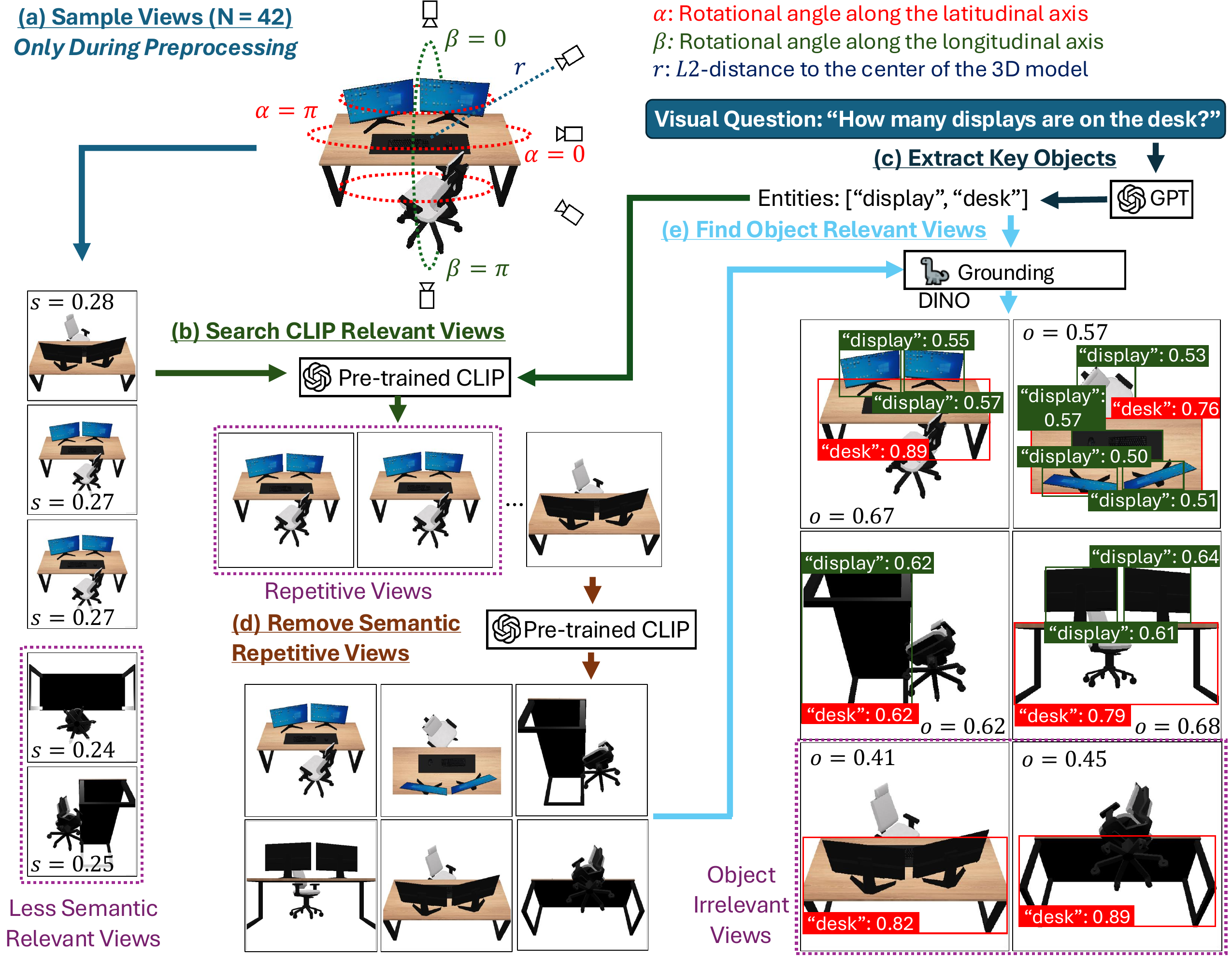}
    \caption{Pipeline for view sampling and selections. $42$ views are first sampled by navigating viewing camera, where $\bm{s}$ refers to the similarity score (a); the VQA pipeline then (c) extracts the key entities from the visual questions, (b) searches CLIP-relevant views and (d) removes semantic repetitive views; (e) the final selected object-relevant views, where $\bm{o}$ indicates object score.}
    \label{fig::pipeline}
\end{figure*}

\vspace{+4px}\noindent{\bf \ul{Stage 1: View Sampling}} 

\noindent When 3D models are initially loaded, \sysname~renders $42$ views using a rotational viewing camera that can potentially cover all details of the target 3D objects.
Each sampled view contains a $512 \times 512$ RGB image ($\bm{I}_i \in \mathbb{N}^{512 \times 512 \times 3}$) and a depth mask ($\bm{D}_i \in \mathbb{R}^{512 \times 512}$), where $i$ indicates the index of each view ($i \in [0, 42)$).
Each perspective is parameterized by $(x, y, z, \alpha, \beta, r)$. $(x, y, z)$ indicates the target center. $\alpha$, $\beta$, and $r$ indicate the rotational angles along the latitudinal and longitudinal axis, and viewing distance~(Figure~\ref{fig::pipeline}a). 
To determine $r$, the diagonal length ($d$) of the bounding box of the 3D model is first computed.
The viewing distance ($r$) - defined as the distance to the center of the 3D model - was sampled from three possible values: $\{0.5d + 0.1, 0.5d + 0.2, 0.5d + 0.5\}$, representing \emph{close}, \emph{medium}, and \emph{far} inspection distances for the viewing camera.
Unlike~\cite{Luo2023, Ge2024VFC}, sampling viewing distance allows both small objects (\eg~the keyboard of a desk model) and large primary objects to be included in the sampled views.
For each possible $r$, we sampled $\alpha$ and $\beta$ to navigate viewing camera observing the 3D model from $14$ different locations: view from top to bottom, bottom to top, and from four locations (\incl~\emph{front-to-back}, \emph{back-to-front}, \emph{left-to-right} and \emph{right-to-left}) on each of three red orbits shown in Figure~\ref{fig::pipeline}a.
While it is possible to exclude less optimal views, such as perspectives from beneath the desk, the goal of this stage is to ensure that all critical visual information is captured in the sampled views.
Subsequent stages will demonstrate how less optimal or irrelevant views can then be filtered out.

\vspace{+4px}\noindent{\bf \ul{Stage 2: View Selection}}

\noindent While it is possible to prompt MLLM with all sampled views alongside the visual question, bad views can confuse MLLM during the answer generation process, causing hallucinations~\cite{Mo2024BridgeQA}.
\sysname~integrates a view selection pipeline (Figure~\ref{fig::pipeline}b - e) to reject bad and irrelevant views.
A \emph{good view} is considered as both \emph{CLIP relevance} and \emph{object relevance}.

\vspace{+4px}\noindent$\bullet$~{\bf Search CLIP-Relevant Views}

\noindent Just as sighted users gather information by exploring pertinent views of the target 3D model, the chosen views should be related to the visual questions.
\sysname~uses CLIP~\cite{CLIP, Radford2021} to identify views that are semantically aligned with the visual question.
This is realized by the \emph{similarity} and the \emph{flatness} score.
A similar caveat in leveraging pre-trained CLIP model to retrieve text-conditioned views of 3D models was used in MemoVis \cite{Chen2024MemoVis}.

\vspace{4px}\noindent{\bf Similarity ($s_i$)}.
The similarity score is computed by the $cosine$ similarities between the CLIP-encoded sampled views and the \emph{comparative prompt}, which is generated by joining key entities of the visual question.
For example, the visual question {``how many displays on the desk?''} yields two entities: {``display''} and {``desk''}, leading to comparative prompt: {``display, desk''}.
With all $s_i$, we use $z$-score filter to reject the views with $z_{\bm{s}, i} = \frac{s_i - \mu_{\bm{s}}}{\sigma_{\bm{s}}}< -1$. 
Views with low similarities indicate low CLIP relevancy.
With the running example and comparative prompt, \sysname~will reject views with $s_i < \mu_{\bm{s}} - \sigma_{\bm{s}} = 0.26$.
Figure~\ref{fig::pipeline}b shows the rejected views with $s_i$ being $0.24$ and $0.25$. 
These rejected views are rendered from the bottom of the desk that are less aligned with the comparative prompt.

\begin{figure*}[t]
    \centering
    \includegraphics[width=\textwidth]{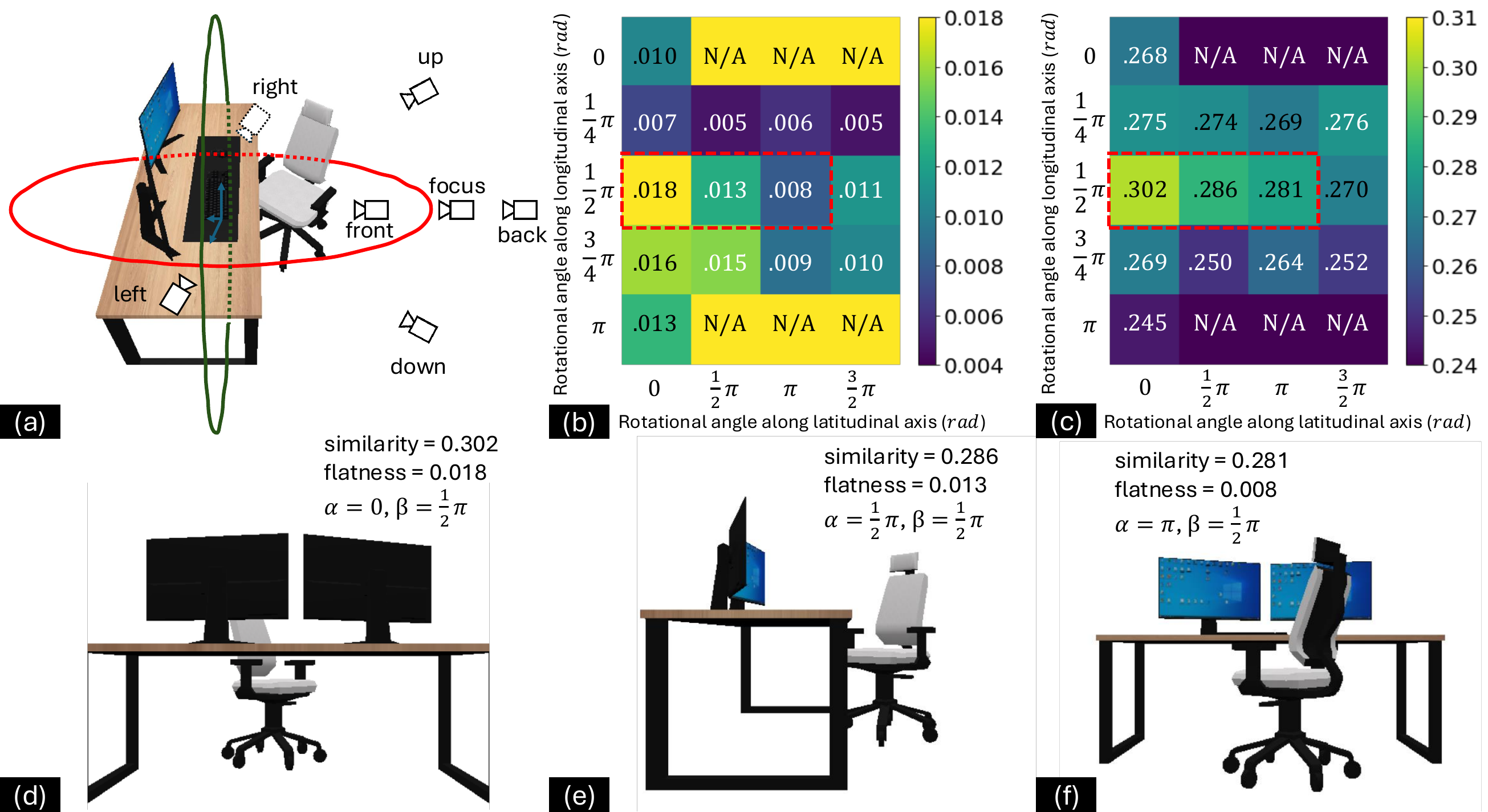}
    \caption{Examples of using flatness score to measure the CLIP relevancy; (a) examples of how flatness score could be approximated; the flatness (b) and similarity score (c) at sampled rotational angles along latitudinal ($\alpha$) and longitudinal ($\beta$) axis, provided $r = 0.5d + 0.2$; (d - e) examples when flatness is used to enhance the reliability of the CLIP relevancy approximations. }
    \label{fig::clip_flat}
\end{figure*}

\vspace{4px}\noindent{\bf Flatness~($f_i$)}.
Views with high $s_i$ \emph{may} suggest strong CLIP relevance, but relying solely on  $s_i$ can be unreliable.
For example, while Figure~\ref{fig::clip_flat}e exhibits a slight higher $s_i$ than Figure~\ref{fig::clip_flat}f 
 ($0.286$ \vs~$0.281$), it would be easier to use Figure~\ref{fig::clip_flat}f for tallying the number of displays.
Therefore, we define the \emph{flatness} ($f_i$) score as the average $L1$-norm of the partial gradient of $s_i$.
Intuitively, $f_i$ measures the difference of $s_i$ between the focused view and the neighboring views.
Provided with a high similarity, the more consistent the score remains with minor adjustments in the viewing camera's perspective, the greater the confidence in the CLIP relevance.
For example, the flatness of the focus view in Figure~\ref{fig::clip_flat}a can be approximated by Equation~\ref{eqn::approx_focus}.
Figure~\ref{fig::clip_flat}\mbox{b - c} show $f_i$ and $s_i$ for all views, provided $r = 0.5d + 0.2$ (\ie~\emph{median} distance between viewing camera and the 3D model).
Figure~\ref{fig::clip_flat}\mbox{d - f} visualize the rendered views, with Figure~\ref{fig::clip_flat}f appearing to be the most suitable for counting the display with the lowest $f_i$.
With $z$-score filter, we reject the views with $z_{\bm{f}, i} = \frac{f_i - \mu_{\bm{f}}}{\sigma_{\bm{f}}}< -1$.

\begin{equation}
\begin{split}
    f_{focus} \approx \frac{1}{6}(|s_{focus} - s_{back}| + |s_{focus} - s_{front}| \\ + |s_{focus} - s_{up}|  + |s_{focus} - s_{down}| \\ + |s_{focus} - s_{right}| + |s_{focus} - s_{left}|)
\end{split}
\label{eqn::approx_focus}
\end{equation}

Finally, we used the image encoder of the CLIP~\cite{CLIP, Radford2021} to remove the views that are well aligned (\ie~ the repetitive views).
Figure~\ref{fig::pipeline}d shows an example in which only one view rendered from the front of the desk is retained.
With the example of Figure~\ref{fig::pipeline}, six CLIP-relevant views are retained.

\vspace{+4px}\noindent$\bullet$~{\bf Search Object-Relevant Views}

\noindent While CLIP relevance narrows down the views, certain views like the bottom-up perspective of the desk model (last column, Figure~\ref{fig::pipeline}d) may fail to address the visual question.
\sysname~uses the Grounding DINO \cite{Liu2023GroundingDINO} to evaluate the \emph{object} score ($o$), measured by the confidence of the existence of key objects.
Incorporating Grounding DINO~\cite{Liu2023GroundingDINO} enables \sysname's pipeline to more explicitly identify and analyze key objects within the sampled views.

To evaluate $o_i$, \sysname~computes the confidence of the bounding box using Grounding DINO~\cite{Liu2023GroundingDINO} by prompting each extracted entity.
$o_i$ is then approximated by averaging all bounding box confidence.
The inferred bounding boxes are omitted if they are perfectly overlapped with the entire 3D model.
A zero-confidence box is added when object recognition fails.
Figure~\ref{fig::pipeline}e shows examples of how the {``\mbox{display}''} and {``\mbox{desk}''} are recognized. 
The two outliers being rendered from the bottom of the desk are rejected with a similar $z$-score filtering. 
Empirically, slightly higher thresholds than the defaults were chosen for box and text in Grounding DINO~\cite{Liu2023GroundingDINO} ($0.50$ and $0.35$).

\vspace{+4px}\noindent{\bf \ul{Stage 3: Answer Generation} }

\noindent With the selected views, \sysname~generates descriptions using both generative- and task-specific recognition-based models. 
%
% how do we synthesize 
We recognize the importance of providing reliable responses for confirmative and informative questions that are associated with factual details, such as the questions inquiring about counting specific objects, the details of the color, and the types of materials (\textbf{DC2}).
%
% why MLLM doesn't work
%
\sysname~first uses LLM to evaluate the type of visual question created. 
Upon \emph{counting}-type questions (\eg~{``how many displays are on the desk''}), \sysname~uses a compositional visual reasoning technique~\cite{Suris2023} to generate the answer with the selected views.
Empirical observation shows that employing compositional visual reasoning yields more accurate answers for counting-type questions with clear task goals.
Without compositional visual reasoning, GPT-4V alone incorrectly considers {``one display''} for Figure~\ref{fig::study_tasks}b.
We use MLLM to fulfill VQA for other types of questions like seeking descriptions of the design style of the desk, as generating accurate answers requires AI to grasp the contextual nuances of entire views, rather than merely focusing on specific components of interest.

\begin{figure*}[t]
    \centering
    \includegraphics[width=\textwidth]{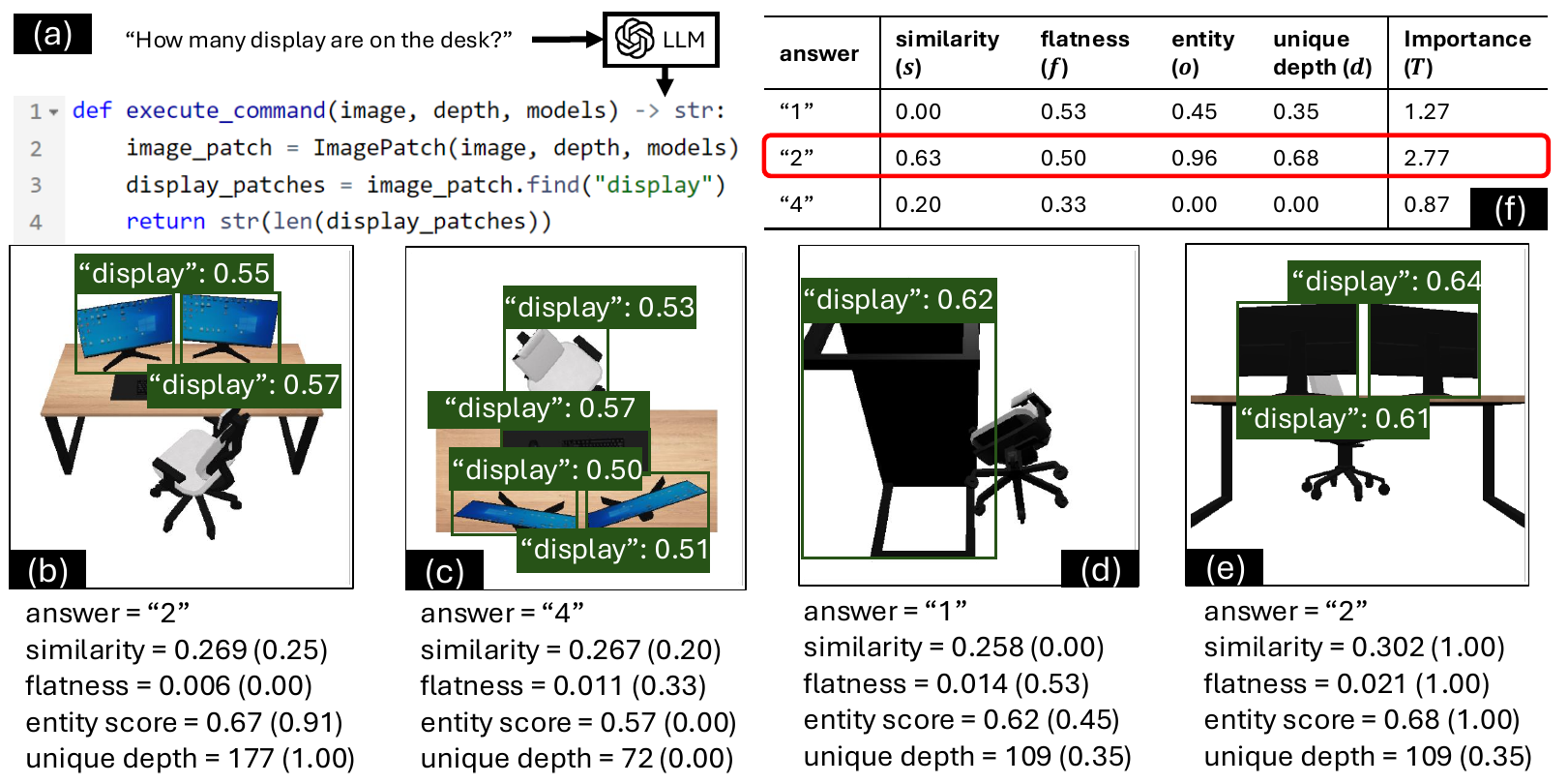}
    \caption{Demonstration of compositional visual reasoning for answer generations in Stage 3. (a) Python code generated by an LLM for compositional visual reasoning; (b - e) recognition results of the ``display'' using Grounding DINO ~\cite{Liu2023GroundingDINO} by evaluating the selected views from Figure~\ref{fig::pipeline}e; (f) synthesis generated answers from selected views.}
    \label{fig::compositional_vqa}
\end{figure*}

\vspace{+4px}
\noindent $\bullet$~{\bf Compositional Visual Reasoning} 

\noindent\sysname~uses LLM to infer the procedural code for generating answers, followed by executing the AI-generated code with recognition-based models.
The prompts used by ViperGPT~\cite{Suris2023} were used for code generation, leveraging the LLM's in-context learning capabilities.
Figure~\ref{fig::compositional_vqa}a shows the generated Python code to count the number of displays.
Instead of using a pre-trained generative model~\cite{Li2022blip, Li2023, openai2024gpt4}, \sysname~leveraged the Grounding DINO ~\cite{Liu2023GroundingDINO} to evaluate each selected view by executing the procedure \texttt{image\_patch.find("display")}.
Figure~\ref{fig::compositional_vqa}\mbox{b - e} visualize the recognition results.

By executing such compositional visual reasoning pipeline, the inferred answers for each selected view might not reach consensus.
\sysname~selects the answer that maximizes the \emph{total importance} ($\hat{k} = \argmax_k(T_k)$).
To compute $T_k$, \sysname~first uses four measures to approximate the importance of the view $i$ associated with each answer: the similarity ($s_i$) and flatness ($f_i$) score computed by CLIP~\cite{CLIP}; the object score ($o_i$) computed by Grounding DINO~\cite{Liu2023GroundingDINO}; and the number of unique depth values computed from the captured depth map ($d_i$, hereinafter referred to as \emph{unique depth}).
The scores for views that result in the same description will be aggregated by averaging.
$T_k$ can then be computed using Equation~\ref{eq::cost}.
Intuitively, a good view for inferring the answers would lead to higher $s_i$, $o_i$, and $d_i$, and lower $f_i$.
$\lambda_s$, $\lambda_o$, $\lambda_d$ and $\lambda_f$ are empirically set to $1$.
Figure~\ref{fig::compositional_vqa}f shows how the answer ``2'' is decided with the largest $T_k$.

\vspace{-.15in}
\begin{equation}\label{eq::cost}
    T_k = \lambda_s s_k + \lambda_o o_k + \lambda_d d_k + \lambda_f (1 - f_k).
\end{equation}
\vspace{-.15in}

\noindent$\bullet$~{\bf VQA by MLLM.}
GPT-4V was used to generate answers by prompting all selected views, along with the textual prompt: {\it ``Given different views of a 3D model. Answer the question in one sentence. Question: \inlinecode{\$\{VQ\}} The answer should be concise''}, where \inlinecode{\$\{VQ\}} is replaced by created visual questions.

Finally, to address \textbf{DC1}, we use a similar approach as \mbox{GenAssist} \cite{Huh2023}, where LLM was used to summarize the similarities and differences with the answers generated based on individual 3D models.
All responses will be automatically updated on the SR accessible table upon completing inferences.

\section{User Studies}\label{sec::evaluation}

Two user studies were conducted.
Study 1 with \emph{BLV users} aims to understand how \sysname~is used to explore 3D models. 
While Study 1 focuses on BLV users' \emph{experience}, asking BLV users to visually assess the accessibility quality of the generated descriptions is difficult.
Similar to \cite{Suhyun2024, Zhang2022}, a second study was conducted with \emph{sighted} users to evaluate the \emph{quality} of the generated descriptions.
To enhance readability, phrases in the generated descriptions were highlighted in \hlgreen{green} for positive comments and in \hlred{red} for negative ones.

\subsection{Study 1: Evaluation of 3D Models Access Experience by BLV Users}\label{sec::study::blv}
Two \textbf{R}esearch \textbf{Q}uestions~(RQs) are focused: {\it how BLV users create visual questions to explore and compare 3D models} (RQ1), and {\it how \sysname's descriptions support them in this process} (RQ2).

\vspace{+4px}\noindent {\bf Participants.}
P1 - P10 (age $M= 33.6$ $SD = 12.4$, \A~\ref{sec::app::blv}) were recruited from three community centers in the Southwestern United States and a Facebook group. 
Low-vision users were included, as they commonly use SRs as well~\cite{ScreenReaders, Szpiro2016}.
All participants identified themselves as experienced SR users who primarily rely on SRs to access visual content on various computing devices.
Among the six participants with partial vision, three do not have usable vision, although they can perceive changes in light.
Participants' experience with SRs ranged from $1$~year to $20$~years.

\vspace{+4px} \noindent {\bf Tasks.}
Each participant was invited to complete three 3D model browsing tasks (T1 - T3).
In each task, participants were instructed to make a purchasing decision by exploring and comparing a set of four 3D models using \sysname.
Participants were instructed to converse with \sysname~to make purchase decisions using the editable table, similar to how they would shop in-person while asking for assistance from the staff. 
Participants can create free-formed visual questions, which may take the form of questions or commands.
The purchase decisions were made when participants informed the researcher and provided justifications. 
The researcher then confirmed the justifications by referring back to the 3D model.
The 3D models used for T1, T2, and T3 include a set of four \emph{pen holders} (T1), \emph{desks} (T2), and \emph{bikes} (T3) (\A~\ref{sec::app::models}).

\begin{figure}[t]
    \centering
    \includegraphics[width=0.48\textwidth]{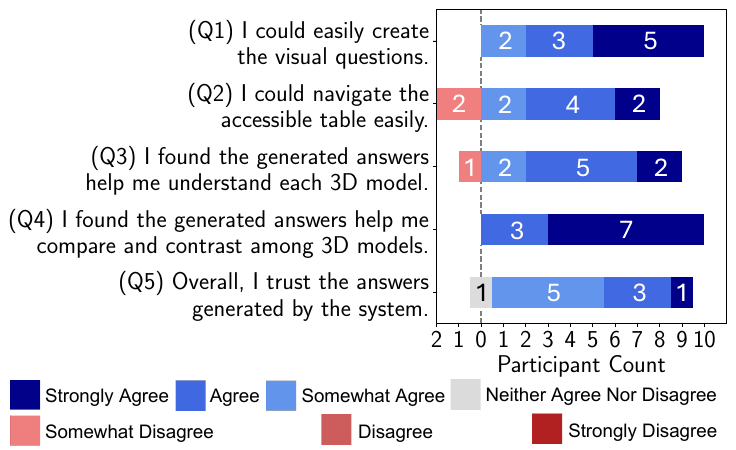}
    \caption{Responses of usability questions of Study 1. Questions were assessed through a 7-point Likert scale.}
    \label{fig::study1_seq}
\end{figure}

\vspace{+4px}\noindent{\bf Procedures and Analysis.}
An online {\it expert review}~\cite{Baecker1995} was conducted with each participant.
After completing the demographic questionnaire, participants were introduced to \sysname and instructed to use T1 to familiarize themselves with it.
Participants then completed the exploration task with T2 and T3 while thinking aloud~\cite{Someren1994}.
The order of tasks was counterbalanced. 
P3 - P7 were instructed to complete T2, followed by T3, whereas others were instructed to complete T3, followed by T2.
%}
%
Finally, participants were invited to complete a survey to evaluate their experience, focusing on visual question creations\footnote{While rating the visual question creation experience, BLV participants were instructed to focus on the cognitive process of coming up with visual questions, disregarding factors related to the ease of typing.} (Q1), navigating the SR-accessible tables (Q2), and the helpfulness of the answers (Q3 - Q5) (Figure \ref{fig::study1_seq}).
The \textbf{A}ccessible \textbf{U}sability \textbf{S}cale (AUS)~\cite{AUSOverview, AUSAnalysis} questionnaire was used to assess usability (Appendix~\ref{sec::app::aus}).
An interview was finally conducted based on participants' responses.
Thematic analysis~\cite{Braun2012, Lazar2017} and deductive and inductive coding approach~\cite{Lazar2017} were used to analyze the qualitative data.
All studies have been approved by the IRB~(see \A~\ref{sec::app::irb} for more details).

\vspace{+4px}\noindent {\bf Results.}
Most BLV participants found \sysname~easy to use, with the overall mean AUS score being $79.5$ ($SD = 14.4$).
The mean AUS scores among JAWS, NVDA and VoiceOver users are $81.9$ ($SD = 4.3$, $N = 4$), $71.3$ ($SD = 33.6$, $N = 2$), and $81.3$ ($SD = 13.0$, $N = 4$)\footnote{The average AUS score for the desktop SRs measured by Fable~\cite{AUSAnalysis} is around $55$.}.
Figure~\ref{fig::study1_seq} shows survey responses.

\vspace{+4px}\noindent $\bullet$~{\bf How BLV users create visual questions? (RQ1)}

\noindent 
All participants were able to create and type visual questions using their chosen screen readers.
The created visual queries include $59.1\%$ informative questions, $13.1\%$ confirmative questions, and $27.3\%$ commands.
Each typed query contained an average of six words ($SD = 1.88$ words).
On average, participants created $3.4$ queries ($SD = 1.1$ queries) using their chosen SR before finalizing the purchase decision.
Participants spent on average $41.3$~s ($SD = 90.4$~s) typing each query.

Most of the participants appreciated the ability to ask questions and remarked it as {\it ``easy''} (P4, P6), {\it ``simple''} (P9), and {\it ``straightforward''} (P10).
Specifically, many participants valued the flexibility of asking customized questions based on their experiences, \eg~{\it ``I like the way I could ask questions, because that way you don't have to have these different questions and information that you might not care about. So, you create your own questions. [...] For example, what if we just wanted to know what color it is, instead of going through a bunch of questions and irrelevant descriptions, you just ask the question''}~(P1).
P2 appreciated how the editable table allowed him to ask follow-up visual questions: {\it ``Sometimes, it will give me the answer like `it is brown'. But how `brown' it is. So I would probably continue to ask it in some other way. With the description, I would also ask to describe more details.''}
Similarly, while choosing a bike model, P7 explicitly mentioned the need for further details: {\it ``Some description says it's a cruiser. Another one says it's a mountain bike. Two of them say it's a BMX bike. I think this one (Figure~\ref{fig::study_tasks}f) for sure, but I would definitely need more detail.''}~
A few participants hinted at the value of asking questions related to the fine-grained details of the 3D models. For example {\it ``what I might ask is something like do the drawers have knobs or handles''}~(P3), {\it ``it is useful to ask about specific details of the 3D model like the handles''}~(P6).
The ability to prompt follow-up questions also helped P2 verify the accuracy of their questions: {\it ``When I hear the description of the bikes, I will look in that description whether it says green or not. So that way, I can make sure that is right. [...] The idea would be to ask somebody, like `Hey, is this really green?' The information might be redundant, but I don't want to disregard it either''}~(P2).

However, a few participants emphasized the need to retain the overview description before VQA.
P4 suggested: {\it ``If (the SR) could read a general overview of each 3D model, that would be nice to have. I could get a complete overview beforehand. Then it would be easier for me to ask questions.''}~
Specifically, P4 initially faced challenges in creating visual questions until she browsed the answers of the initial overview-type question: {\it ``What does the desk look like?''} 
From a different perspective, P3 suggested: {\it ``It would be beneficial to have questions that might have more useful answers in a pool. If a person asks a question, it could kind of match it to questions in that pool. When other users have asked a similar question, it could just use the questions in that pool. Because, sometimes, it's all about the wording.''}

\vspace{+4px}\noindent $\bullet$~{\bf How the generated answers can help BLV users access 3D models? (RQ2)}

\noindent Most participants positively rated the helpfulness of the \sysname-generated answers (Figure~\ref{fig::study1_seq}, Q3).
Only P4 gave a negative rating, but our interview revealed that the rating was due to their limited table navigation experience with SRs. 
P4 was still positive about the generated answers: {\it ``[The answers] are very cool and useful.''}~
Other participants considered the generated answers to be {\it ``informative''} (P3, P8), {\it ``detailed''} (P1) yet {\it ``not too lengthy''} (P2, P5, P9). 
P2 emphasized: {\it ``I feel the answers are as succinct as it can be. It is not lengthy. It's just trying to provide the answers as thoroughly as possible.''}~

Most participants can {\it ``picture the 3D models in [the] mind''}~(P6).
For example, P1 drew an analogy to his in-person shopping experience: {\it ``Exploring the answers can help me picture how the desk looks like. It's just like when I go to the store, I could talk to sellers and feel the different desks.''}
Some participants liked the value of having descriptions focused on the key components of the 3D models. For example: {\it ``This is also a very great description about the desk and the attached bookshelves''}~(P1) and {\it ``That's useful. It tells me the key things, it says an office and chair setup''}~(P2).
P9 liked the extra details beyond the focus of the questions: {\it ``I like the fact that it gives more details. For example, it says it is a mountain bike [...] which would also be best suited for a city.''}~
Likewise, P10 and P3 mentioned that additional fine-grained details about the bike models (\eg~{\it ``chunky tires''}) and the desk models (\eg~{\it ``spacious storage space''}) were useful in helping them make purchase decisions.

\vspace{+4px}\noindent $\bullet$~{\bf How the summaries of similarities and differences help BLV users compare and contrast 3D models? (RQ2)}

\noindent All participants either \emph{agreed} or \emph{strongly agreed} to the usefulness of the summarized similarities and differences.
Example testimonies include: {\it ``Having the comparisons there in the table really does make things a lot more streamlined''} (P3) and {\it ``The similarities and the differences are quite good and give nice descriptions. [...] It helps me a lot on making the purchasing decision''} (P5).
%
% read compare first, and then individual model.
Many participants prefer to consume summaries of similarities and differences, followed by browsing the descriptions of individual 3D models.
For example: {\it ``I love to hear similarities and differences first, it makes me better understand what to focus while hearing the descriptions for individual models''}~(P6) and {\it ``The summary of differences is quite helpful. I don’t need to navigate back to the answers for each model''}~(P1).

% read individual model first, and then compare and contrast
Contrarily, a few participants preferred to browse the answers for each 3D model and use the summary of compare and contrast as a confirmation.
For example, {\it ``I would like to see all four answers first that describe each 3D model so that I can also make my own comparisons as to what the system is telling me. Then, I'll read the compare and contrast to kind of confirm what I was thinking before''}~(P7).
Beyond confirmation, some participants used the summaries to assist in making their final purchase decisions: {\it ``the similarities and the differences are quite good and give nice descriptions.[...] It helps me a lot in making the purchasing decision''}~(P5).

\begin{figure}[t]
    \centering
    \includegraphics[width=0.48\textwidth]{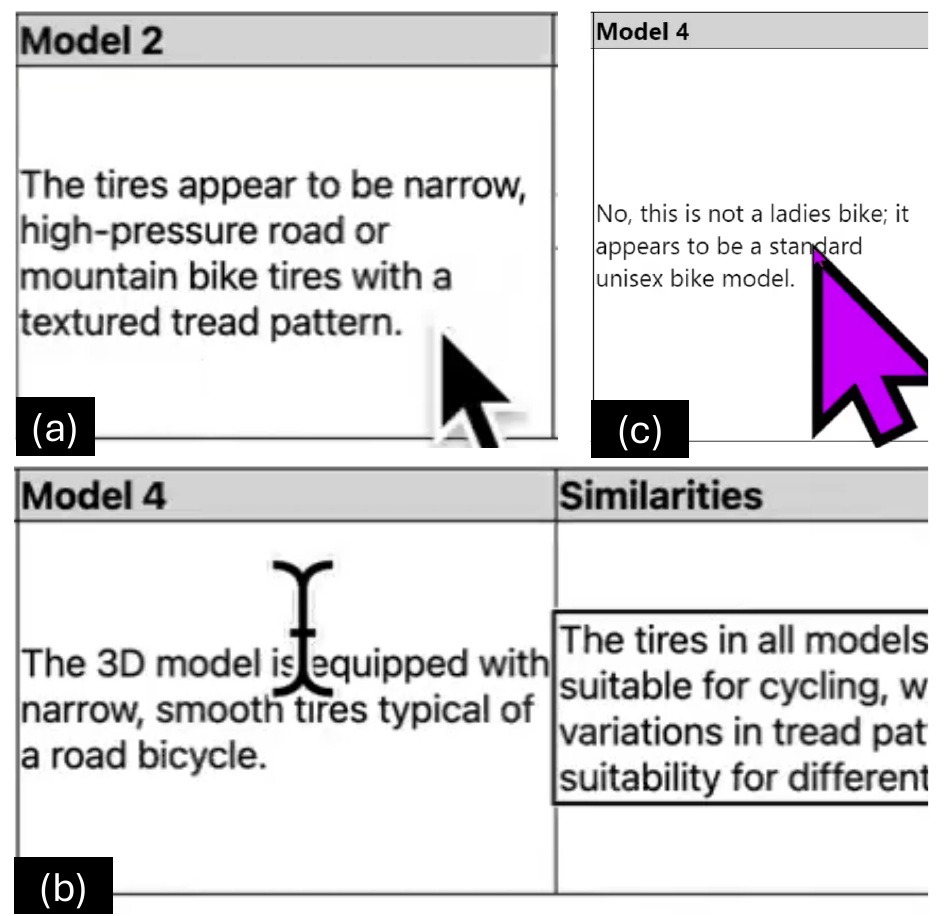}
    \caption{Examples of how low vision users navigate the table; (a - b) P5 leveraged the change of mouse cursor to locate its position; (c) P9 was trying to use the mouse cursor with high contrast color to locate the focused cursor.}
    \label{fig::study1_example_sr_bigarrow}
\end{figure}

\vspace{+4px}\noindent$\bullet$~{\bf \emph{Perceived} quality of the generated descriptions (RQ2).}
\noindent All participants successfully made the instructed purchasing decisions with correct justifications. 
Precisely evaluating BLV users' perception of the generated descriptions' quality is impractical due to their limited visual capabilities.
Yet, among $264$ generated descriptions, $17$ ($6.44\%$) descriptions were explicitly noted by participants as confusing, lacking critical details or less useful during the think-aloud process\footnote{It is impractical for BLV users to note \emph{all} misleading descriptions due to their limited visual abilities.
\S\ref{sec::study::accuracy} discusses the in-depth evaluation of the \emph{quality} of the generated descriptions by \emph{sighted} users with prior \alt~text experience.}.
Specifically, four ($1.52\%$) descriptions were noted as confusing \emph{and} misleading.
For instance, the description of {\it ``It is a \hlred{computer table}''} was synthesized for \mbox{Figure} \ref{fig::study_tasks}a when P1 asked, {\it ``Is it a computer table?''}~ 
P1 identified the inaccuracies with the answer of the follow-up visual questions: {\it ``It is a small \hlgreen{office cabinet}, about the size that fits a laptop and some books.''}~
Grounded on this generated description, P1 commented: {\it ``So it's a cabinet. I thought it is a regular office desk.''}
Nine ($3.41\%$) descriptions were commented for missing critical details, which were later clarified through descriptions from follow-up visual questions.
For example, P7 initially failed to notice the white bookshelf in Figure~\ref{fig::study_tasks}c upon hearing the description: {\it ``The desk is brown.''}
This detail becomes clearer upon hearing:
{\it ``the desk has a brown wood color with a \hlgreen{white bookshelf} [...]''}, synthesized from a follow-up question.

Similarly, P2 understood the {\it ``L-shape''} after browsing the description in the follow-up overview question: {\it ``[...] a white vertical support on one side [...]''} while exploring Figure~\ref{fig::study_tasks}c.
Finally, four~($1.52\%$) descriptions were noted for the less helpfulness of numerical information for mentally visualizing the 3D models.
For example, P7 appreciated the usefulness of the description, {\it ``The desk is around \hlred{75 cm} tall,''} ~although she admitted that {\it ``the number doesn't help her mentally picture the desk.''}

\vspace{+4px}\noindent $\bullet$~{\bf How do BLV users navigate the editable and SR-accessible tables? (RQ2)}

\noindent 
On average, participants spent $58.6$~s ($SD = 46.7$~s) browsing and consuming the descriptions generated by \sysname\footnote{In a real-world, ecologically valid setting, the actual time for browsing and consuming \sysname-generated descriptions may be shorter than our measurement, since participants were asked to think aloud while completing the tasks. For example, most participants tended to repeat and verbalize their understanding of the 3D models while browsing the table.
}.
Eight participants believed that table navigation is easy with the help of their SRs (Figure~\ref{fig::study1_seq}, Q2).
For example, {\it ``Navigate the table is very easy with the screen reader shortcuts. I am sure most screen reader users should know how to use these shortcuts''}~(P1).
P2, P3, and P5 favored the table design that allows the focused cursor to be navigated with two DOF (\textbf{D}egrees \textbf{o}f \textbf{F}reedom) with SRs' browse mode~\cite{BrowseFocusMode}.
This was evidenced by the observation that P2 and P5 tended to review all previously generated descriptions in the same column after consuming each newly generated one.
Example testimonies include: {\it ``If I go up and down, I could compare the different answers for the same bike model. But if I go left and right, I could compare the different bike models for the same question. That's really good! Table is actually the perfect way to convey this information''}~(P2) and {\it ``It's very easy to go back through the history and get answers for each desk that the user is looking at''}~(P3).
P10 valued the opportunities provided by the table to skip irrelevant information: {\it ``It would be a nightmare if you put all descriptions together in a paragraph because it would just read and read and read. And that just sounds awful to me. Whereas, with a table, you can move through it and see different cell content [...] I like to have some sort of interactions to navigate among cells.''}

\vspace{+4px}\noindent{\bf Inexperienced SR users.}
Although the table provides two DOF for navigating descriptions with SRs, few participants believed that {\it ``it might be challenging for those with less table navigation experience''}~(P4).
This may partially explain P4’s outstanding browsing time of $119.5$~s ($SD = 39.0$~s), which was $103.9\%$ higher than the overall average.
P4 was also frequently observed moving the focused cursor outside of the \sysname~interface, \eg~to elements on the web browser or desktop.
While P7 believed that using a table is a better design than organizing information into different level of headings (\eg~{\it ``It's very easy to navigate as everything is all there, so I don't have to keep scrolling by heading''}), P4 contrasted this opinion due to limited experience of table navigation: {\it ``Screen reader is very linear. It would be nice if things could be designed as kinda straight down. It just makes things a little easier [...] having headings can make us jump to them easier.''}
Similarly, P2 suggested to add a dedicated shortcut to announce guides for those with limited table navigation experience.

\vspace{+4px}\noindent{\bf Low-vision users.}
%
% While users without usable vision relied on the keyboard, few low-vision users leveraged their residual vision to navigate the focused cursor. 
Few low-vision users leveraged their residual vision to navigate the focused cursor by enlarging the mouse pointer (Figure~\ref{fig::study1_example_sr_bigarrow}) and/or configuring it to be a high-contrast color (Figure~\ref{fig::study1_example_sr_bigarrow}c).
%
%Without central vision, P5 leveraged the change\footnote{The peripheral vision is highly sensitive to motion~\cite{Post1986}.} of the mouse cursor (Figure~\ref{fig::study1_example_sr_bigarrow}a - b) to locate its position: {\it ``I make my mouse cursor big. I could still see the location of the mouse. I could just use the mouse to help me navigate the cell, and the screen readers could read the content for me. It is a bit faster and more intuitive''}.
%
Without central vision, P5 leveraged the change of the mouse cursor (Figure~\ref{fig::study1_example_sr_bigarrow}a - b) to locate its position: {\it ``I make my mouse cursor big. I could still see the location of the mouse. I could just use the mouse to help me navigate the cell, and the screen readers could read the content for me. It is a bit faster and more intuitive.''}~
P9 used a high-contrast pointer to locate the cell and hear the text content (Figure~\ref{fig::study1_example_sr_bigarrow}c).
While low-vision users might prefer accessing texts and images visually~\cite{Szpiro2016}, none of the low-vision users directly interacted with the 3D models (Figure~\ref{fig::accessible_table}d).

\subsection{Study 2: Evaluation of the Quality of Generated Descriptions}\label{sec::study::accuracy}
\emph{Visually} validating the generated descriptions is impractical for BLV users. 
Although Section~\ref{sec::study::blv} reported the \emph{perceived} quality of the generated descriptions by BLV users based on qualitative data, a \emph{online survey study} was conducted with \emph{sighted} users experienced in writing and evaluating \alt~texts to assess the \emph{quality} of \sysname-generated descriptions.
While existing AI research (\eg~\citealp{Banerjee2005Meteor, Nils2019SBertScore, Zhang2020BertScore, Antol2015, Manas2024, Papineni2002BLEU}) introduced different accuracy scores for automatic VQA evaluations, applying these methods to evaluate \sysname~remains challenging. 
\First, existing benchmark datasets are often created by sighted users (\eg~\citealp{Goyal2017, Antol2015}) and/or designed for images (\eg~\citealp{VizWizDataset}), which differs from \sysname's use cases.
Although datasets~\eg~ScanQA~\cite{Azuma2022ScanQA, Dai2017} may be used as benchmarks for 3D VQA tasks, their focus on large-scale scanned environments for spatial understanding differs from ours -- emphasizing the accessibility of simpler 3D models for BLV users.
Finding existing datasets with visual questions and reference descriptions created by BLV users for 3D models is difficult.
Therefore, human evaluation was employed and carried out with sighted participants.
\Second, while accuracy is one crucial metric for evaluating the performance of VQA pipelines~\cite{Antol2015}, evaluating \emph{quality} focuses on how the generated descriptions help BLV users access 3D models via SRs by addressing visual questions - going beyond mere accuracy.

\vspace{+4px}\noindent {\bf Participants and Evaluation Dataset}.
$30$ \emph{sighted} participants were recruited with experience of writing and evaluating~\alt~text as the human evaluators (E1 - E30, age $M = 25.2$, $SD = 3.4$, \incl~$22$ males and eight females).
Questions from Study 1 were collected and used to generate answers by two additional \emph{baseline conditions}, driven by the \emph{image}-based VQA pipelines: $C_{2D + GenAssist}$ and $C_{2D + MLLM}$.
\sysname~can directly work with 3D models, whereas our baseline conditions only accept single-image input.
Hence, for baseline conditions, we used the \textit{canonical views} of the 3D models and fed them as the image input. 
These canonical views serve as effective default perspectives for 3D models. 
Specifically, we downloaded these 3D models from CGTrader~\cite{cgtrader} and used the perspective of the corresponding thumbnail specified by the assets creators as the canonical views.
The three conditions include:

\begin{itemize}[noitemsep, topsep=2px, leftmargin=*]

\item {$C_{2D + GenAssist}$}: The canonical view was used as the input image (Figure~\ref{fig::canonicalviews} in \A~\ref{sec::app::models}). GenAssist~\cite{Huh2023}, which leverages the BLIP 2 ~\cite{Li2023}, was used to generate the answers to the visual questions collected from Study 1.

\item {$C_{2D + MLLM}$}: Similar to $C_{2D + GenAssist}$, the same canonical view was used as the input. GPT-4V~\cite{openai2024gpt4} was used to generate answers. Unlike BLIP 2~\cite{Li2023}, GPT-4V exhibits an improvement in benchmarks of common-sense questions~\cite{Li2024}.

\item {$C_{\sysname}$}: The same \sysname-generated descriptions collected from Study 1.

\end{itemize}
\vspace{+2px}

\vspace{+4px}\noindent {\bf Procedures.}
With the aforementioned approach, $264$~samples were created for evaluation.
Each sample includes a visual question, three descriptions generated under the aforementioned three conditions, and the associated 3D model or all four 3D models\footnote{As \sysname~generates descriptions to assist BLV users in accessing individual 3D models and comparing them, evaluating each generated description may involve using a single 3D model or all four 3D models.}.
After completing demographic questions, for each sample, participants were instructed to explore 3D models using a 3D viewer and a video showcasing various perspectives.
Participants were then asked to rate each of three descriptions, similar to~\cite{Suhyun2024}, focusing on six criteria: {\it accuracy}, {\it clarity}, {\it informativeness}, {\it understandability}, {\it length appropriateness}, and {\it preference}, using a $7$-point Likert scale~(Appendix~\ref{sec::app::quality}).
An optional text field was provided for subjective rationales.
For each sample, the order of the three descriptions was randomized; sighted participants were unaware of how the descriptions were generated. 
All evaluation samples were randomized and split into $30$ surveys, distributed to different participants.
On average, each survey took $43.4$~min ($SD = 10$~min).

\begin{figure}[t]
    \centering
    \includegraphics[width=.48\textwidth]{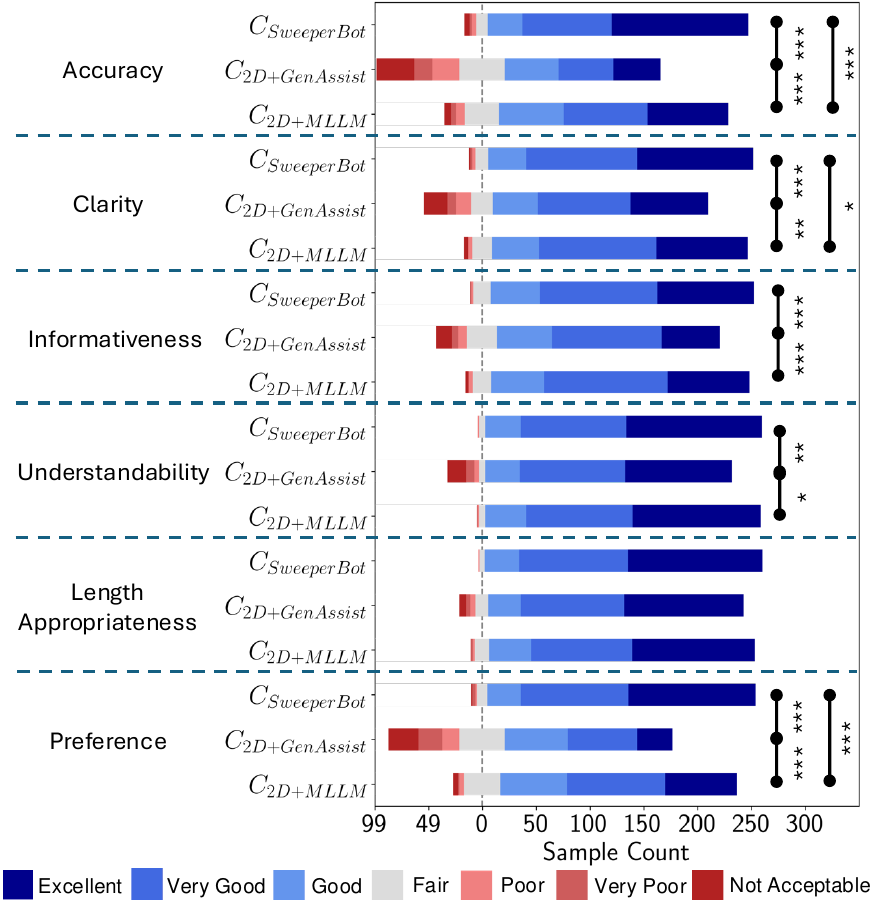}
    \caption{Survey responses evaluating the quality of the descriptions using a 7-point Likert scale (* $\bm{ = p < .05}$, ** $\bm{ = p < .01}$, *** $\bm{ = p < .001}$).}
    \label{fig::study2_rating}
\end{figure}

\vspace{+4px}\noindent {\bf Analysis.}
Kruskal‐Wallis Test~\cite{Kruskal1952} was used to analyze the collected responses ($\alpha = 0.5$).
Dunn's test \cite{Dunn1964} with Holm–Bonferroni adjustment \cite{Holm1979} was used for the post-hoc test.
$\eta^2$ was reported to evaluate the effect size. 
The empirical thresholds $.01$, $.06$ and $.14$ of $\eta^2$ were used for \emph{small}, \emph{moderate} and \emph{large} effect size~\cite{Tomczak2014}.
Thematic analysis~\cite{Braun2012} and the inductive coding approach~\cite{Lazar2017} were applied to analyze $104$~textual comments.
\A~\ref{sec::app::codebook} shows final codebook.

\vspace{+4px}\noindent {\bf Results.} 
Overall, the statistical significance was found in terms of accuracy ($H_2 = 124.13$, $p < .001$, $\eta^2 = .155$), clarity ($H_2 = 30.26$, $p < .001$, $\eta^2 = .036$), informativeness ($H_2 = 27.65$, $p < .001$, $\eta^2 = .033$), understandability ($H_2 = 12.55$, $p = .002$, $\eta^2 = .013$), and preference ($H_2 = 143.69$, $p < .001$, $\eta^2 = .018$) (Figure~\ref{fig::study2_rating}).
{\it Even without canonical views - a common scenario in existing 3D applications and emerging text-to-3D GenAI services - we found that the quality of the descriptions generated by \sysname~is \textbf{comparable} to that of the baselines reliant on canonical views.}

\vspace{+4px}\noindent $\bullet$~{\bf Inferring readable and understandable answers.}
Although most descriptions generated by $C_{\sysname}$ and $C_{2D + MLLM}$ are understandable, participants noted the unreadability of some descriptions inferred by $C_{2D + GenAssist}$, potentially leading to a more negative rating regarding \emph{informativeness} and \emph{understandability} (Figure~\ref{fig::study2_rating}).
For example, E26 noted: {\it ``The last description seems to be weird. Is it a typo?''}~ for a $C_{2D + GenAssist}$-generated answer: {\it ``The seat and handlebars are made of carbon fiber. \hlred{The seat and handlebars are made of carbon fiber. The seat and handlebars are made of carbon fiber. The seat and handlebars}''}~ (Figure~\ref{fig::study_tasks}f) regarding the visual question: {\it ``Describe the seats and handles?''}~ created by P5.
Contrarily, \sysname~generated a clear response: {\it ``The bike has a padded black seat and black handlebars with no visible grips.''}
A few incomplete answers were also noted.
For example, E1 remarked, {\it ``incomplete and could be misleading''} in response to the $C_{2D + GenAssist}$-generated answer: {\it ``The bike is yellow and has a black frame. It has a front wheel and a rear wheel. It has a front fork and a rear fork. \hlred{It has a front brake and a}''} (Figure~\ref{fig::study_tasks}e), regarding the visual question {\it ``Describe the bike in more detail''} created by P6.
In contrast, \sysname~generated: {\it ``The bike features a yellow frame with black accents, a front suspension fork, a flat handlebar, a black saddle, and chunky tires, suggesting it's designed for off-road use.''}~
Finally, participants also mentioned the unclear pronoun observed in the $C_{2D + GenAssist}$-generated answers.
For example, E21 commented: {\it ``What does `this' mean?''} for the $C_{2D + GenAssist}$-generated answer: {\it ``The desk looks like \hlred{this},''} for the visual question {\it ``What does the desk look like?''} created by P4.
Differently, \sysname~describes: {\it ``The desk has a modern design with a dark wood finish, featuring a white raised shelf with different items and a globe on the top''} (Figure~\ref{fig::study_tasks}c).

\vspace{+4px}\noindent $\bullet$~{\bf Coverage of specific key object(s).}~ 
Participants highlighted multiple examples when baseline conditions failed to capture the small yet critical objects, of the 3D model. 
This possibly cause the significantly higher rating regarding \emph{accuracy}, \emph{clarity} and \emph{preference} for $C_{\sysname}$~ (Figure~\ref{fig::study2_rating}).
For example, E19 noted: {\it ``There is no tablet on the desk,''} in response to a $C_{2D + MLLM}$-generated answers to a visual question created by P4 - {\it ``What does the desk look like?''} 
This testimony can be verified by comparing $C_{2D + MLLM}$-generated answer: {\it ``The desk appears to be a compact, modern-styled cabinet desk with closed doors, topped with a lamp, \hlred{a digital tablet}, and a coffee cup,''} with the $C_{\sysname}$-inferred answer: {\it ``The desk is a rectangular cabinet-style desk with a flat top, on which there are various items including \hlgreen{a laptop}, a lamp, \hlgreen{books}, and a coffee cup,''}~ while inspecting the 3D model shown in Figure~\ref{fig::accessible_table}a.
Similarly, E7 criticized: {\it ``The desk is not white, the door is white,''} for the $C_{2D + GenAssist}$-generated answer: {\it ``The color of the desk is white''} for Figure~\ref{fig::study_tasks}a. 
This merits of covering key object(s) in \sysname~can potentially help mitigate the setbacks related to misleading descriptions, noted by P1 and P7.

\vspace{+4px}\noindent $\bullet$~{\bf Extra fine-grained details could be beneficial.}~
Many participants praised the additional information to explain the answer to a moderate extent.
For example, E10 preferred the additional detail generated by $C_{\sysname}$ ({\it ``No, this is not a ladies bike. It appears to be a unisex bike \hlgreen{without the traditionally lower crossbar associated with ladies' bikes}''}) \vs~$C_{2D+MLLM}$ ({\it ``No, it is not designed specifically for ladies. It's a unisex design''}), because {\it ``the mentioning of the `lower crossbar' is useful''}.
This finding was validated by BLV participants in Study 1; for example, P10 found the additional descriptions of the bike tires helpful.
Despite expressing the same meaning, E13 favored the descriptions generated by $C_{\sysname}$ ({\it ``The desk is of a modern, minimalist style''}) and $C_{2D+MLLM}$ ({\it ``The desk is a modern, minimalist style with a clean design and simple lines''}), over $C_{2D+GenAssist}$ ({\it ``Modern''}).

\section{Discussion}\label{sec::discussion}
\subsection{Practical Implications}\label{sec::discuss::3da11y}
\sysname~showed a novel approach to assisting BLV users in accessing 3D models through automatic camera view analysis and VQA.
While BLV users may use existing \alt~text of the canonical view(s) and the companion textual profile to access 3D models (\eg~FP1), \sysname ~ showed how to make 3D models accessible even \emph{without} canonical view(s) and profile descriptions.
This is important as canonical view(s) and textual descriptions might not always be available, particularly for the emergent GenAI-generated 3D models~\cite{3dfyai}.
This section discusses five practical implications.

\vspace{+4px}\noindent{\bf Preserving existing \alt~texts}.
Contributions around \sysname~do not suggest removing the existing \alt~text.
Study 1 underscored the recommendations (\eg~from P4) to preserve the existing \alt~texts, allowing for an efficient \emph{overview} for each 3D model.
Although BLV users can still query the overview descriptions of the 3D models, retaining the existing \alt~text may better streamline the workflow of accessing 3D models.
This design conforms to the \emph{visual information-seeking principle}~\cite{Shneiderman1996} and auditory information-seeking principle~\cite{Zhao2004} that emphasize the critical role of \emph{overview} during the visual information seeking workflow for sighted users.

\vspace{4px}\noindent{\bf Assisting visual question creations}.
\sysname~may necessitate designing features to assist BLV users in formulating, articulating, and typing visual questions.
While P3 suggested recommending potentially useful questions for BLV users, we highlighted that visual question creation is also related to participants' specific needs and experience.
Future design may explore novel techniques to help BLV users in creating visual questions, by contextualizing on the specific 3D applications and previously asked questions.

\vspace{4px}\noindent{\bf Emphasizing on the key objects.}
Study 1 revealed the importance of describing key components of the 3D model in aiding BLV users to mentally visualize the model, such as the tires of the bike and the objects on the desk.
We observed that several key phrases associated with specific objects typically played a significant role in helping BLV participants make their final purchasing decisions.
While \sysname's view selection demonstrated the combined use of CLIP~\cite{CLIP} and Grounding DINO~\cite{Liu2023GroundingDINO} in curating relevant views, the current approach to generating answers lacks explicit control for producing more focused answers.
Future design may explore ways to generate more efficient and succinct descriptions centered around key objects.

\vspace{+4px}\noindent{\bf Awareness of AI hallucinations. }
Despite the design of view selections and the capabilities of allowing BLV users to converse with \sysname, both BLV and sighted participants pointed out that some responses from \sysname~may occasionally be misleading or inaccurate.
Nevertheless, Study 1 unveiled that BLV participants can still identify the possible misleading responses using follow-up visual questions. 
Following this observation, future work may explore techniques, \eg~providing more detailed rationales for generated responses, to help BLV users recognize inaccurate AI responses.

\vspace{4px}\noindent{\bf Supporting users with limited SR experience}.
Although most participants acknowledged the advantages of two DOF while navigating the table using SRs, we suggested the importance of accommodating low-vision users and those with limited SR-based table navigation experience:
BLV users with limited SR-based table navigation experience might prefer the \sysname-generated answers to be organized in a heading-based information structure;
Low vision users might prefer to use a customized mouse cursor (and/or other low vision aids~\cite{Szpiro2016}) to locate the cell, leading to a more efficient table navigation.
While we showed the VQA's potential for BLV users accessing 3D models, advancing \sysname's capabilities necessitates further investigation into alternative methods for presenting answer layouts.

\vspace{4px}\noindent{\bf \sysname~helps with accessing 3D models, but it might not be a panacea}.
Despite showing the promising of \sysname, it does not imply that the \sysname~will be the best solution to help BLV users access 3D models.
While appraising the usefulness of the generated descriptions, P2 commented: {\it ``If I could physically use it, I would still want to physically touch it. [...] I would love to sit on a chair in front of the desk to feel what it's like to work at this desk''}.
Future work may explore how \sysname~can complement alternative methods (\eg~\citealp{Siu2019ShapeCAD}) to improve the visual accessibility of 3D models, grounded on broader applications.

\subsection{Applicability of \sysname} \label{sec::discuss::general3d}

While focusing on \emph{simple} 3D models that \emph{only} need to be explored \emph{externally}, the ideas of \sysname~can be integrated into many existing 3D applications and used for broader 3D browsing tasks.

\vspace{+4px}\noindent{\bf Integrate with existing 3D applications.}
Although the preloaded high-fidelity 3D models was used, \sysname~can be integrated into existing web-based 3D model repositories and computer-aided design tools as a plugin, allowing BLV users to efficiently search and browse 3D models on demand~\cite{autodesk3dmaxint}.
While FP2 currently relies on ChatGPT and sighted friends, we expect that the 3D model repository with \sysname~can help FP2 achieve the same goal with significantly less time and mental effort. 
\sysname~may also be integrated into existing e-commerce sites with 3D preview features, \eg~IKEA~\cite{IkeaApp}, where the seller-provided textual profiles and canonical views may not cover key details.

\vspace{+4px}\noindent{\bf Generalize to broader 3D tasks.}
While \sysname~was evaluated in the context of browsing 3D models for online shopping, the core ideas can be generalized to other 3D tasks. 
For instance, \sysname~can enhance the accessibility of today's 3D printing workflows by enabling BLV users to efficiently explore and compare 3D models in online repositories.
This \sysname-powered workflow can potentially address FP1's 3D printing barriers when browsing models from online repositories while avoiding high cost and bulky setups~\cite{Siu2018ShapeShift, Siu2019ShapeCAD}.
\sysname's VQA pipeline alone may help BLV users better understand specific 3D models.
For instance, \sysname~can be used in mechanics education to help BLV students understand mechanical designs as effectively as sighted students.
\sysname's VQA pipeline can also support \emph{sighted} users in exploring 3D models. 
By providing textual references, it simplifies the process of visually capturing key characteristics of the 3D models, reducing the need for tedious navigation through various perspectives using a keyboard and mouse.
In addition, \sysname~can make the emergent GenAI-powered 3D design tools more accessible to BLV users.
While tools like meshy.ai\footnote{Meshy.ai: \url{https://www.meshy.ai}. Accessed on July 9, 2025} allow users to easily create 3D models with textual prompts; it is challenging for BLV users to comprehend and distinguish the subtleties among the generated models due to the lack of creator-selected canonical views and textual descriptions.
\sysname~may be integrated with these kinds of GenAI-driven 3D design tools to help BLV users explore and compare AI-synthesized 3D models.
Finally, we envision that VQA enhanced by view analysis can be extended and applied to existing AI-powered scene description systems.  
For instance, by integrating with today's sensing system (\eg~\cite{Agarwal2019, Boovaraghavan2023}), \sysname's VQA pipeline can be integrated into existing mixed reality instruction systems (\eg~\cite{Chen2023PaperToPlace, Nguyen2025PaperToPlace}) to help BLV users complete fine-grained tasks.

\subsection{Limitations}\label{sec::discussion::limitation}

Grounded in our studies, the limitations of \sysname~can be summarized in fourfold.

\vspace{+4px}\noindent {\bf Latency.}
\sysname~takes around $30$ seconds to generate the answers, constrained by the latency of GPT-4V~\cite{openai2024gpt4}.
The length limit of the CLIP's context window may cause inference failure upon an unusually long input visual question \cite{CLIP, clipvitb32}.
More recent and future MLLM like GPT-5 \cite{GPT5} might mitigate these known limitations.

\vspace{+4px}\noindent {\bf Participants and real-world deployment.}
\First, the formative study included only two participants due to the practical challenges of recruiting individuals with both SR and 3D experience.
While generated design considerations have been successfully demonstrated during our design process and validated through two user studies, future work may include participants with broader 3D-related experience beyond just 3D printing; for example, the formative study may recruit participants with non-congenital blindness who acquired 3D experience prior to losing their sight.
Further design iterations may be conducted based on the findings reported in \S~\ref{sec::evaluation}.
\Second, \sysname~was evaluated by $10$ BLV users with varying levels of SR experience in a controlled setting. 
Although participants have different SR experience (in terms of years of experience, types of blindness, SRs being used, and the ways to use SRs), future work can focus on SR users with a broader range of visual impairments.
For example, by deploying \sysname~in real internet applications, researchers may conduct ecologically valid evaluations of the experience and quality of the generated descriptions.

\vspace{+4px}\noindent {\bf 3D models.}
\sysname~\emph{only} focuses on simple 3D models that only need to be explored \emph{externally}.
The ideas of view sampling and selections based on visual questions can be extended to larger scenes like a 3D environments. 
To support accessing 3D models at this scale, future work can explore methods for sampling views that can selectively cover and analyze only critical elements within the 3D scene.
For example, \sysname~might leverage the mesh structure of the 3D model or recent SAM3D \cite{Yang2023SAM3D} to understand the primary key objects in the scene. 

\vspace{+4px}\noindent {\bf Evaluating the quality of \sysname-generated descriptions.} 
The quality of the answers was evaluated by sighted users with experience in creating and evaluating \alt texts. 
While Study 1 discussed BLV participants' perception of the quality of generated descriptions using qualitative data, future work may expand Study 2 by involving BLV users.
Although it is valid for sighted evaluators to assess the quality of the AI-generated descriptions \cite{Gleason2020TwitterA11y, Suhyun2024, Zhang2022}, future research may compare the mental models of BLV users with and without the use of \sysname~using a pretest-posttest study \cite{Lazar2017}.
Second, Study 2 only evaluated the quality of \sysname-generated descriptions by comparing them with two existing image-based VQA pipelines. 
Further evaluations may be conducted by comparing \sysname-generated descriptions to \alt~ text descriptions created by professional \alt~ text writers.
Finally, we only evaluated $264$ samples using the visual questions generated from Study 1.
It is crucial to recognize the necessity for more extensive evaluations of \sysname~on large-scale curated benchmarks.

\section{Conclusion}\label{sec::conclusion}
\sysname~showed the feasibility of using VQA to assist SR users in exploring and comparing 3D models - the critical and indispensable foundational task for many 3D applications.
An expert review with $10$ BLV user experienced in SRs demonstrated how \sysname~can support BLV users accessing and comparing 3D models.
The quality of the generated descriptions was evaluated by a second survey study with $30$ sighted participants.

%TC:ignore
\begin{acks}
We thank the insightful feedback from our colleagues at Adobe Research, University of California San Diego, Florida International University, and the anonymous reviewers from the International Journal of Human-Computer Interaction (IJHCI).
We are grateful to the Blind Community Center of San Diego, San Diego Center for the Blind, Blind Center of Nevada, and Nicholas Ho for helping us with advertising and assistance with participant recruitment.
\end{acks}

\balance
\bibliographystyle{ACM-Reference-Format}
\bibliography{reference}

\newpage
\appendix
\section{Ethical Disclaimer}\label{sec::app::irb}
All studies have been approved by the \textbf{I}nstitutional \textbf{R}eview \textbf{B}oard~(IRB).
All \textbf{P}ersonal \textbf{I}dentifiable \textbf{I}nformation (PII) has been removed.
As required by IRB, we have obtained participants' consent on collected data through web-based forms, video, audio and screen recordings.
For the Study 1 with BLV users, participants were rewarded with $\$30$ Amazon gift card upon the successful completion of the study.
While no monetary incentives were awarded in other user studies, participants were introduced further on our projects and more general AI and assistive technologies.

\section{Demographics of the Blind and Low Vision (BLV) Participants}\label{sec::app::blv}

\begin{figure*}[h!]
    \centering
    \includegraphics[width=\textwidth]{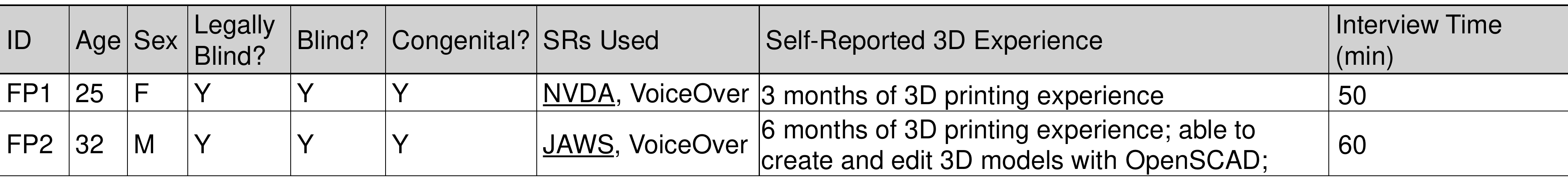}
    \caption{Demographics of formative study participants. The primary SRs being used are underlined. ``SR YoE'' refers to \textbf{Y}ear of \textbf{E}xperience of using SRs as the \emph{primary} tool for accessing visual content on computing devices.}
    \label{fig::formative_blv_participants}
\end{figure*}

\begin{figure*}[h!]
    \centering
    \includegraphics[width=\textwidth]{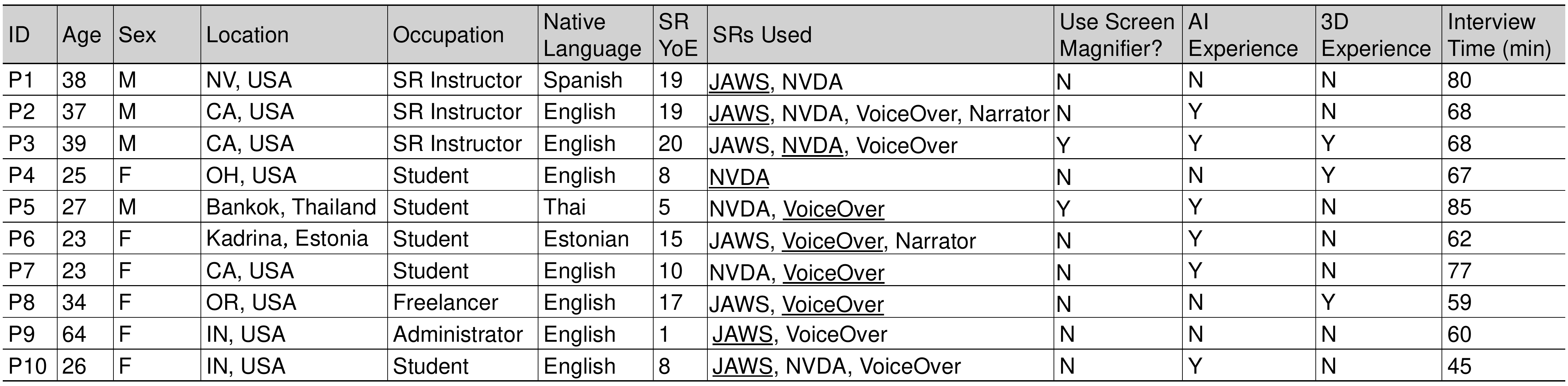}
    \caption{Demographics of 10 BLV participants. SRs being used throughout the study are underlined. ``SR YoE'' refers to \textbf{Y}ear of \textbf{E}xperience of using SRs as the \emph{primary} tool for accessing visual content on computing devices.}
    \label{fig::blv_participants}
\end{figure*}

\begin{figure*}[h!]
    \centering
    \includegraphics[width=\textwidth]{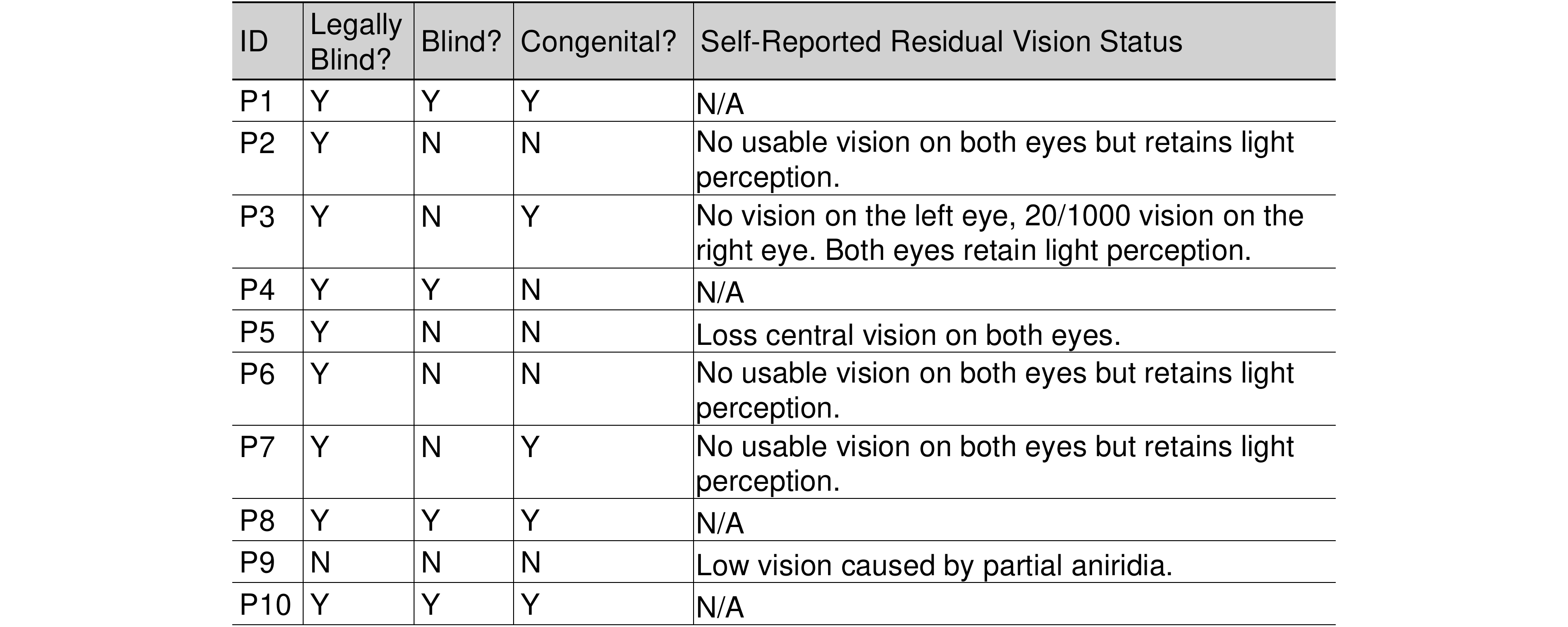}
    \caption{Participants' self-report BLV conditions. ``Legally blind'' describes the BLV condition where visual acuity falls below the threshold defined as legally visually impaired for legal purposes. ``Blind'' indicates that the BLV participants cannot see or detect light. ``Congenital'' refers to visual impairment present at birth in BLV participants.}
    \label{fig::blv_participants_vision}
\end{figure*}

Figure~\ref{fig::formative_blv_participants} shows the demographics of two recruited blind participants for the formative study~(Section~\ref{sec::formative}).
Figure~\ref{fig::blv_participants} shows the demographics of $10$~ recruited BLV participants for Study 1~(Section~\ref{sec::study::blv}).
The self-reported BLV conditions of the recruited participants can be referred to in Figure~\ref{fig::blv_participants_vision}.
Participants from the formative study were \emph{excluded} from Study 1.
All recruited BLV participants have proficient typing skills.

\clearpage
\section{3D Models for User Study}\label{sec::app::models}
\begin{figure*}[h!]
    \centering
    \includegraphics[width=\textwidth]{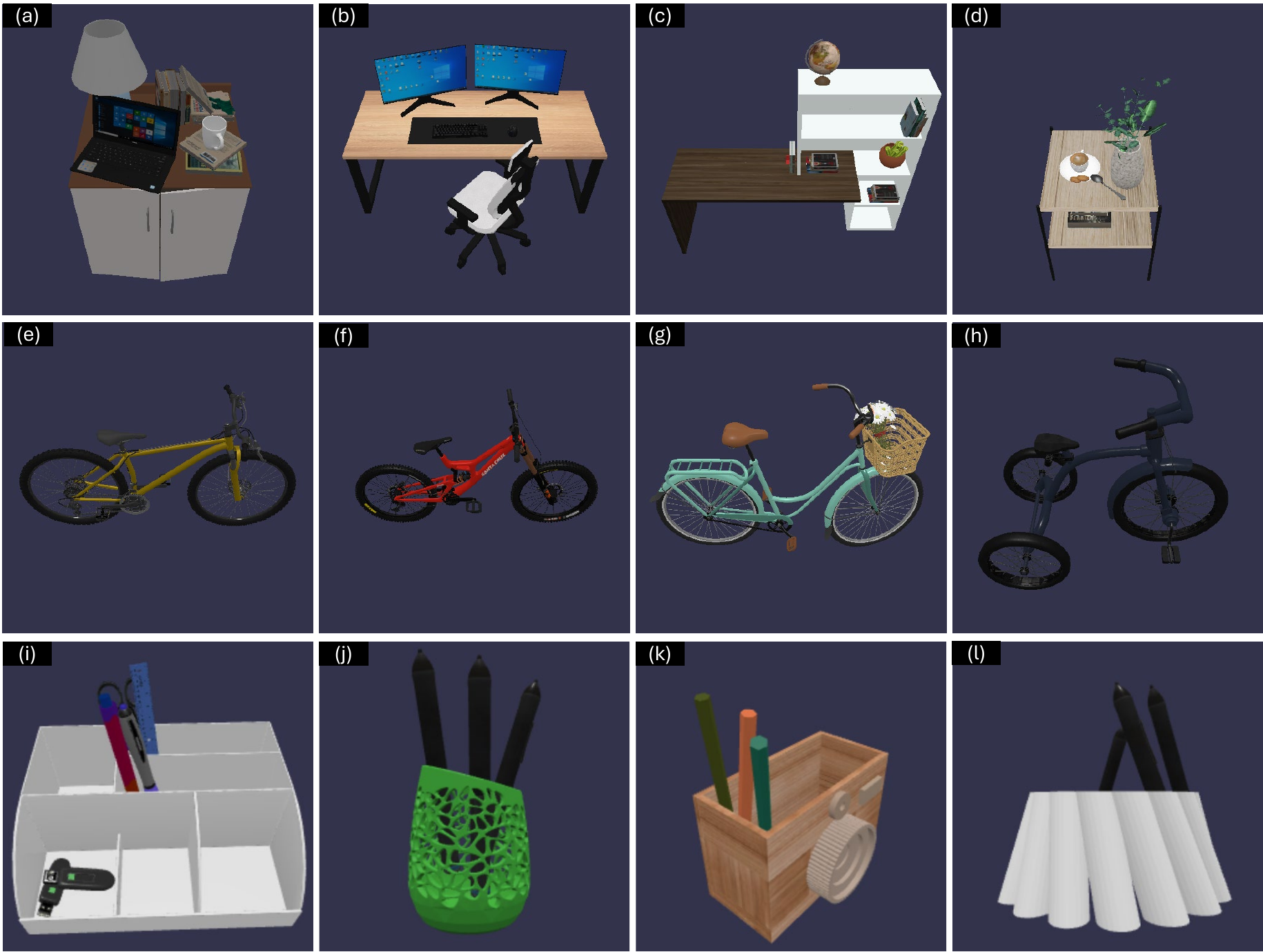}
    \caption{3D models used for user studies; \mbox{(a - d)} four desk models; \mbox{(e - h)} four bike models; \mbox{(i - l)} four pen holder models. 3D models \mbox{(a - h)} were used in the final user studies, whereas the models \mbox{(i - l)} were used for BLV participants to learn and get familiar with using \sysname. }
    \label{fig::study_tasks}
\end{figure*}

\begin{figure*}[h!]
    \centering
    \includegraphics[width=\textwidth]{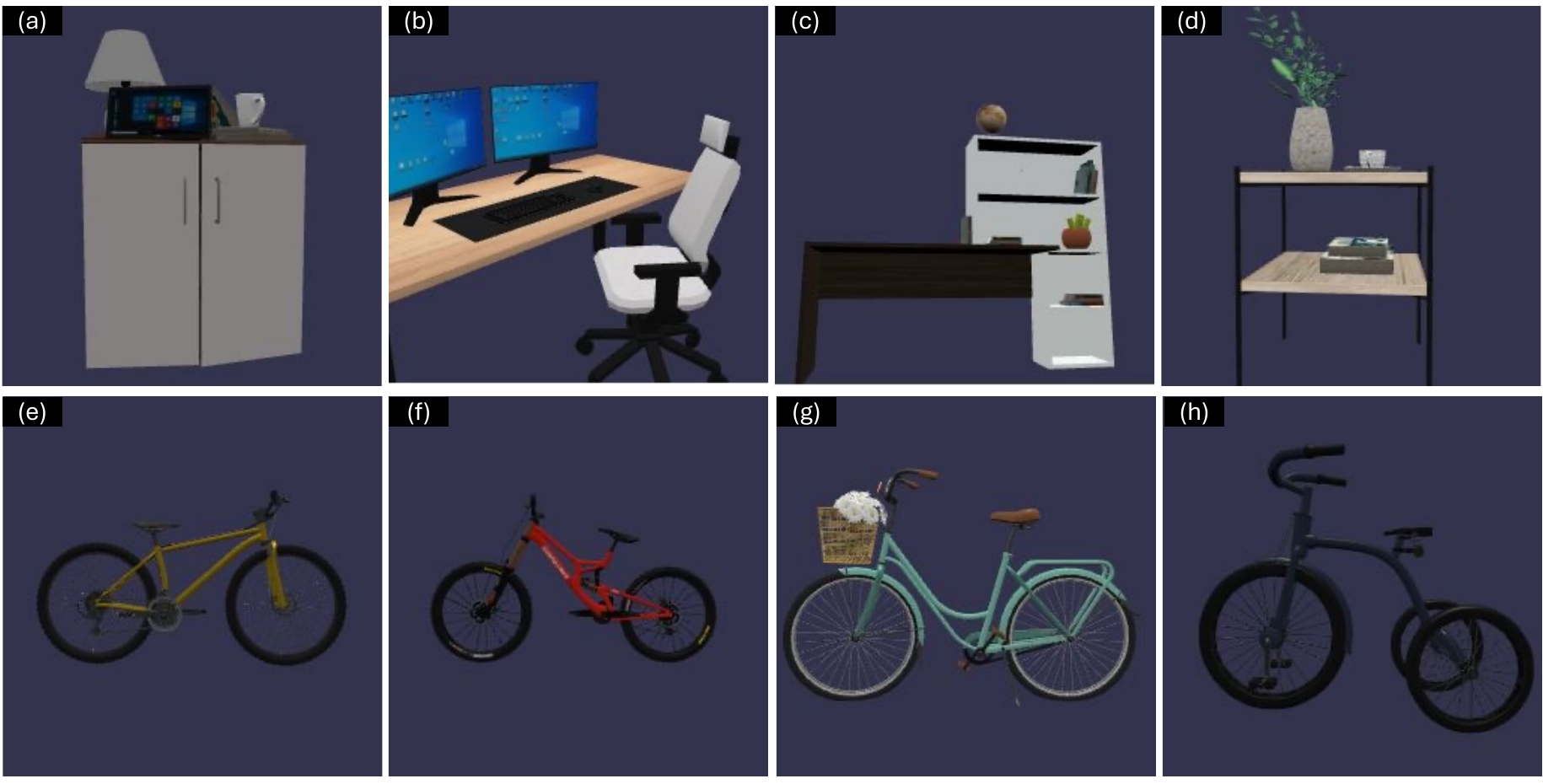}
    \caption{Canonical views used for study 2 (Section~\ref{sec::study::accuracy}); (a - d) four desk models; (e - h) four bike models.}
    \label{fig::canonicalviews}
\end{figure*}

Figure~\ref{fig::study_tasks} shows three sets of 3D models for user studies (\S~\ref{sec::evaluation}).
In particular, the 3D models of the pen holders in Figure~\ref{fig::study_tasks}i - l were only used by the participants to familiarize themselves with \sysname.
Figure~\ref{fig::canonicalviews} shows the canonical views that we used for the second study~(Section~\ref{sec::study::accuracy}).
3D models of \emph{desks} and \emph{bikes} were used for user studies (\S~\ref{sec::evaluation}), as they are common objects that most BLV participants can relate to and have experience purchasing.
Although both T2 and T3 were used for assessing \sysname, completing T2 requires participants to access various key objects, such as books and displays, whereas the completion of T3 requires participants to focus more on the style and aesthetic of the 3D models. 
In addition, 3D models that are less closely aligned with their respective catalogs were intentionally included. 
These 3D models were used to simulate the online shopping experience, where search results may be less relevant to users' queries. 
For instance, a cabinet (Figure~\ref{fig::study_tasks}a) is listed under the ``desk'' catalog.

\clearpage
\section{Codebook and Themes from Qualitative Data Analysis}\label{sec::app::codebook}

\begin{figure*}[t]
    \centering
    \includegraphics[width=\textwidth]{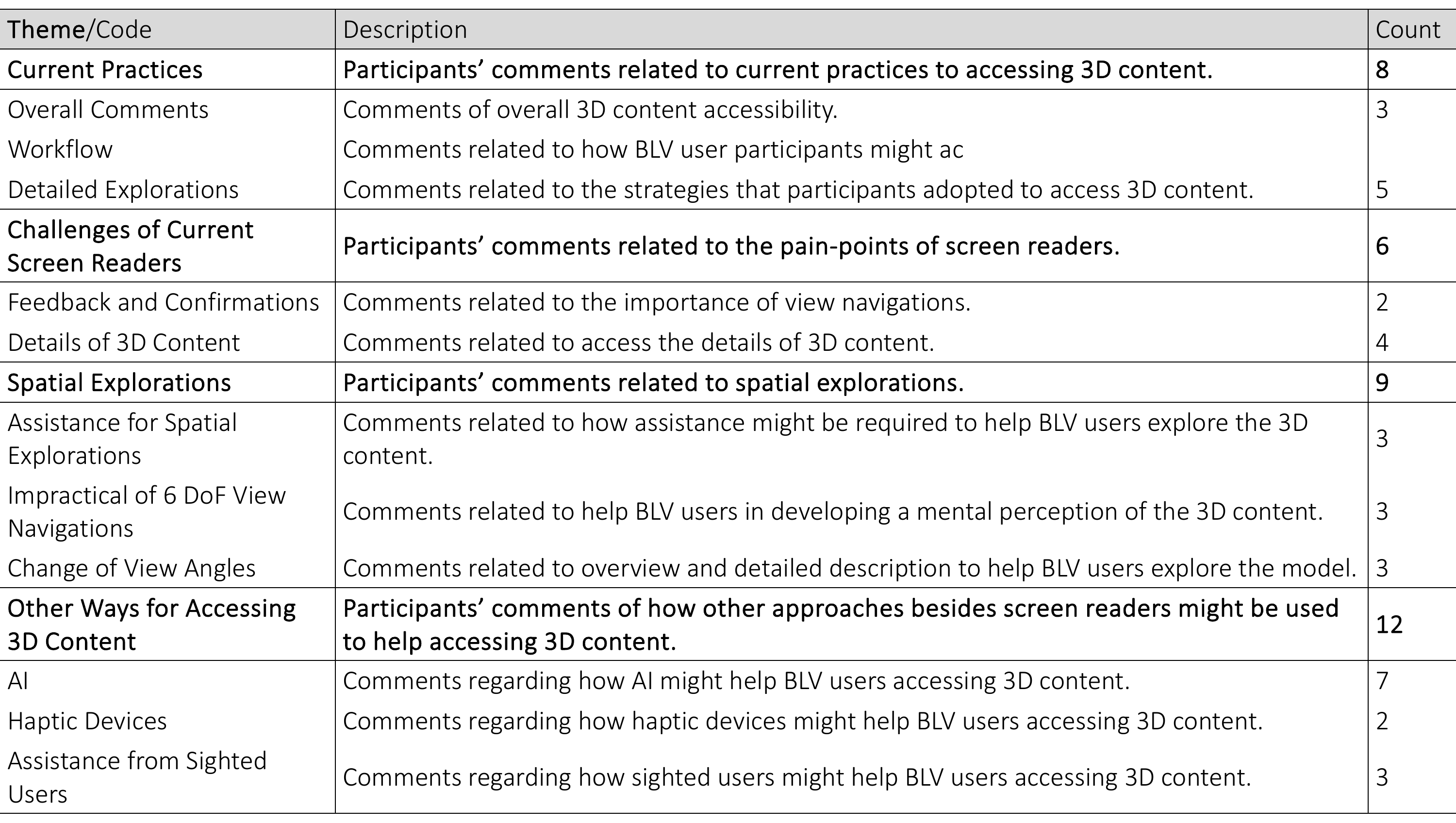}
    \caption{The codebook that resulted from our qualitative analysis of the formative study. ``Count'' refers to the number of quotes for each theme or code. Multiple codes may be assigned to one quote.}
    \label{fig::codebook_formative}
\end{figure*}

\begin{figure*}[t]
    \centering
    \includegraphics[width=\textwidth]{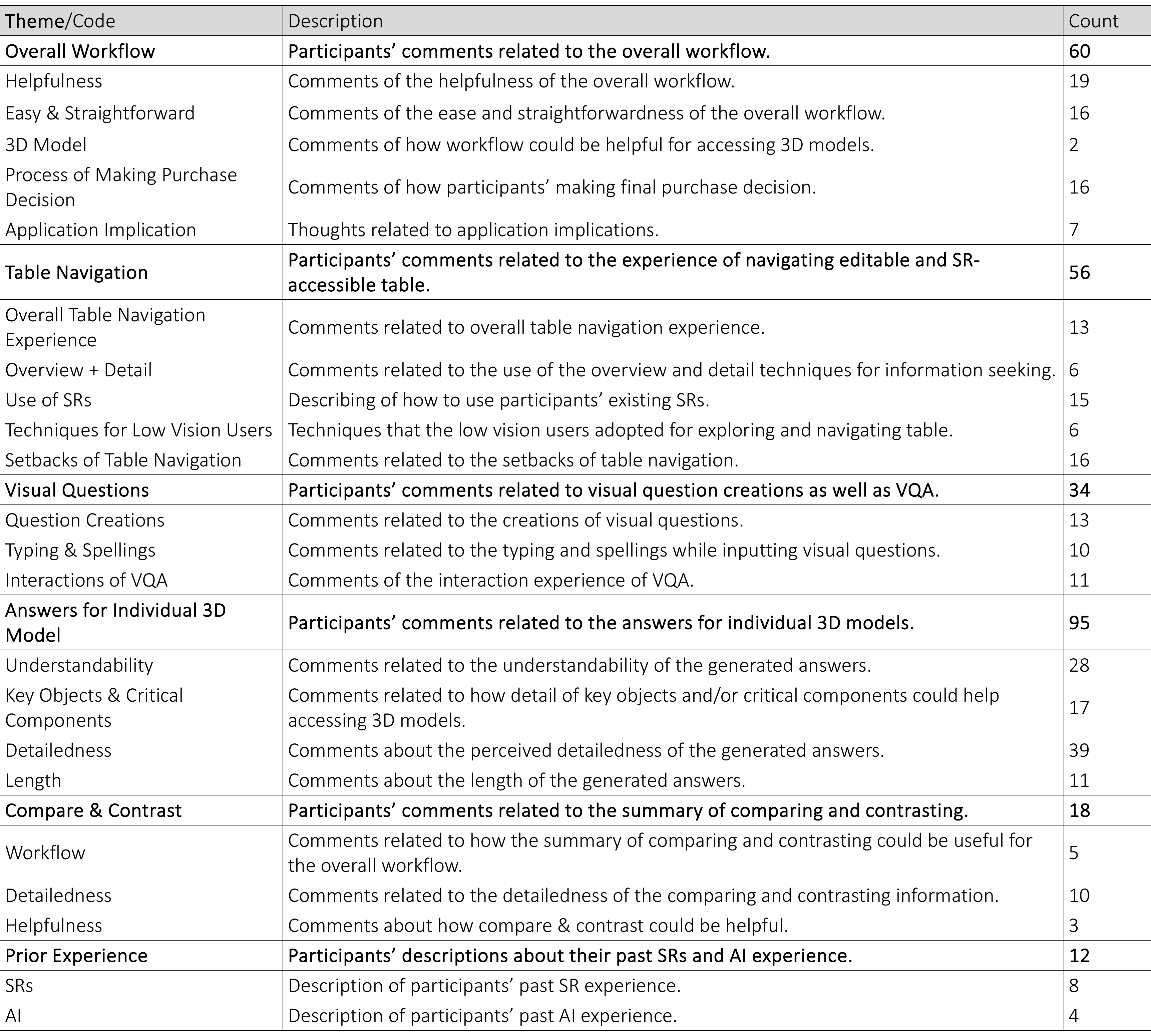}
    \caption{The codebook that resulted from our qualitative analysis of Study 1. ``Count'' refers to the number of quotes for each theme or code. Multiple codes may be assigned to one quote.}
    \label{fig::codebook_study1}
\end{figure*}

\begin{figure*}[t]
    \centering
    \includegraphics[width=\textwidth]{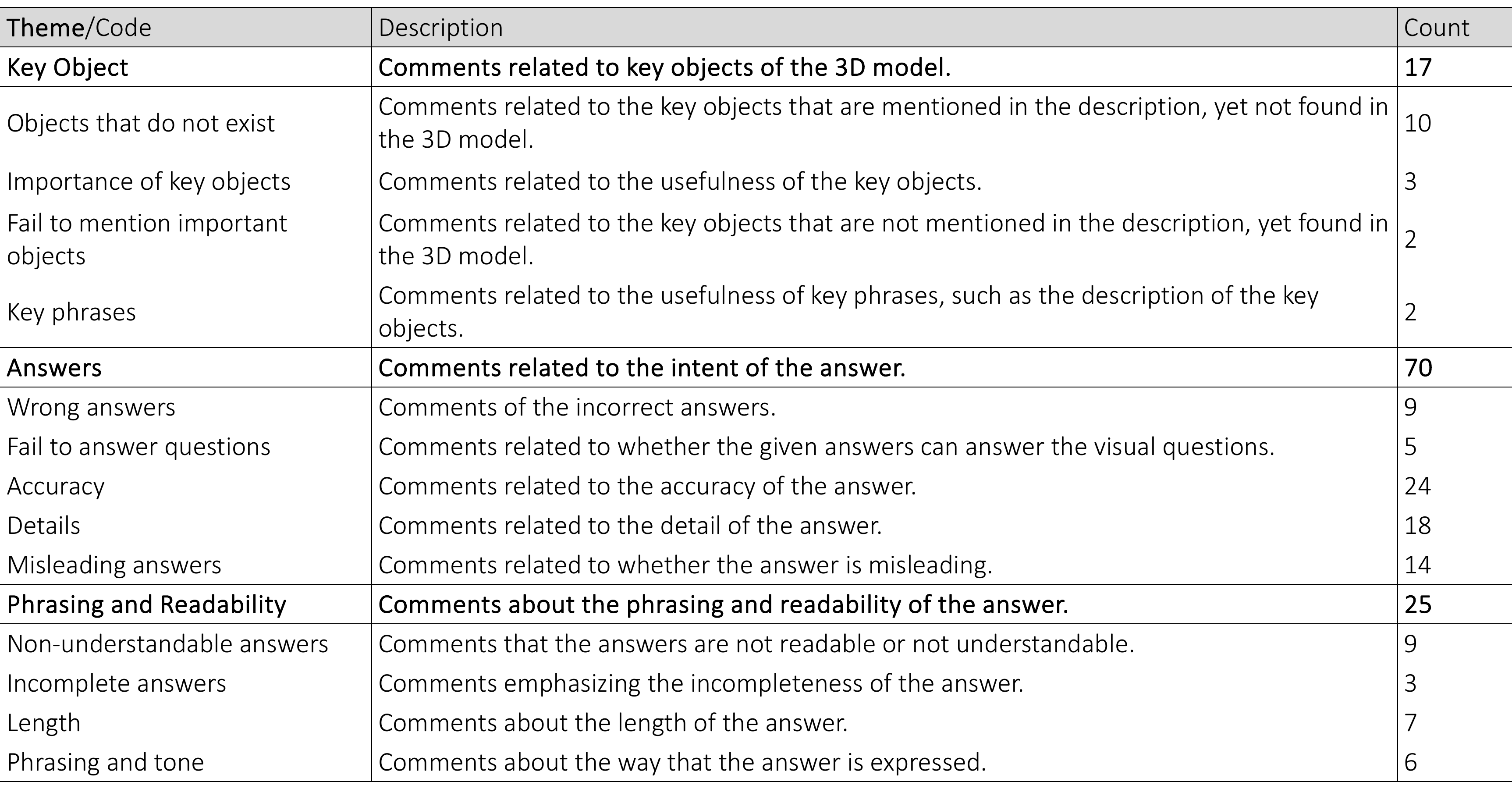}
    \caption{The codebook that resulted from our qualitative analysis of Study 2. ``Count'' refers to the number of quotes for each theme or code. Multiple codes may be assigned to one quote.}
    \label{fig::codebook_study2}
\end{figure*}

This section provides supplementary material for the themes and codebooks resulting from our qualitative data analysis.
Figure~\ref{fig::codebook_formative} shows the themes and codebooks during the analysis of the formative study (Section~\ref{sec::formative}).
Figure~\ref{fig::codebook_study1} and Figure~\ref{fig::codebook_study2} demonstrate the themes and codebooks obtained from the analysis of Study 1 and Study 2, respectively~(Section~\ref{sec::study::blv}).

\clearpage
\section{Accessible Usability Scale Questionnaire}\label{sec::app::aus}

This section provides supplementary material of the \textbf{A}ccessible \textbf{U}sability \textbf{S}cale questionaire revised from \citep{AUSOverview, AUSAnalysis}, which was used in the first user study (Section~\ref{sec::study::blv}).
Participants were required to rate how strongly they agree with the following statement in a $5$-point Likert scale.

\begin{itemize}[noitemsep, topsep=0pt, leftmargin=*]

  \item {\bf AUS1}:~{``I think that I would like to use this system frequently to access 3D model.''}

  \item {\bf AUS2}:~{``I found the system unnecessarily complex.''}
  
  \item {\bf AUS3}:~{``I thought the system was easy to use.''}
  
  \item {\bf AUS4}:~{``I think that I would need the support of a technical person to be able to use this system.''}
  
  \item {\bf AUS5}:~{``I found the various functions in this system were well integrated.''}
  
  \item {\bf AUS6}:~{``I thought there was too much inconsistency in this system.''}
  
  \item {\bf AUS7}:~{``I would imagine that most blind and low vision users with screen reader experience would learn to use this system very quickly.''}
  
  \item {\bf AUS8}:~{``I found the system very cumbersome to use.''}
  
  \item {\bf AUS9}:~{``I felt very confident using the system.''}
  
  \item {\bf AUS10}:~{``I needed to learn a lot of things before I could get going with this system.''}

\end{itemize}

The final AUS score can be computed by Equation~\ref{eqn::aus}, where the statement indices are used to represent the responses rated by the participants.

\begin{equation}
\begin{split}
AUS = 2.5 \times [(AUS1 + AUS3 + AUS5 + AUS7 + AUS9 - 5) \\ + (25 - AUS2 - AUS4 - AUS6 - AUS8 - AUS10)]
\end{split}
    \label{eqn::aus}
\end{equation}

\section{Questionnaire for Quality Evaluations of VQA Results}\label{sec::app::quality}

This section provides supplementary material of the questions used for evaluating the quality of \sysname-generated answers.
The questionnaire has been used in the second study (Section~\ref{sec::study::accuracy}), and was designed and used in prior accessibility research, \eg~\citep{Suhyun2024, Zhang2022}.
Specifically, participants were instructed to evaluate the quality of the VQA results in terms of six measures, listed below, where the token \inlinecode{\$\{VQ\}} was replaced by the visual question in the sample.

\begin{itemize}[noitemsep, topsep=0pt, leftmargin=*]

\item {\bf Accuracy}: ``Given the 3D model, on a scale of 1 to 7, how accurate the description below can answer the question: \inlinecode{\$\{VQ\}}. 1 means worst. 7 means best.''

\item {\bf Clarity}: ``Given the 3D model, on a scale of 1 to 7, how clearly articulated is the description below for the question: \inlinecode{\$\{VQ\}}. 1 means worst. 7 means best.''

\item {\bf Informativeness}: ``Given the 3D model, on a scale of 1 to 7, how informative the description below can answer the question: \inlinecode{\$\{VQ\}}. 1 means worst. 7 means best.''

\item {\bf Understandability}: ``Given the 3D model, on a scale of 1 to 7, how understandable the description below can answer the question: \inlinecode{\$\{VQ\}}. 1 means worst. 7 means best.''

\item {\bf Length Appropriateness (Length)}: ``Given the 3D model, on a scale of 1 to 7, how appropriate is the length of the description below can answer the question: \inlinecode{\$\{VQ\}}. 1 means worst. 7 means best.''

\item {\bf Preference}: ``Given the 3D model, on a scale of 1 to 7, how much do you prefer the description below that can answer the question: \inlinecode{\$\{VQ\}} 1 means worst. 7 means best.''

\end{itemize}
\section{Implementation}\label{sec::app::implementations}

The front end of \sysname~was implemented as a browser-based application using \texttt{React.js}.
While rendering the table, we used the \texttt{arial-label} tag to ensure SRs always vocalize the model index instead of relying on BLV users' memory.
We designed a Flask-based backend and deployed the required VLFMs and LLMs.
We used the checkpoint of \texttt{ViT-B/32}~\citep{clipvitb32} for CLIP, 
\texttt{gpt-3.5-turbo-1106} as the LLM and \texttt{gpt-4-vision-preview} as the MLLM~\citep{openaiModel}.
The LLM and MLLM we selected were the most recent pre-trained models available to us during the time this work was completed in April 2024.
The concept and design of \sysname~are adaptable to newer LLM and MLLM versions, such as GPT-4.5~\citep{GPT45}, at the time of this manuscript's submission.
For Grounding DINO~\citep{Liu2023GroundingDINO}, we used the weights \texttt{groundingdino\_swint\_ogc.pth}, and \texttt{swin\_T\_224\_1k} as the backbone~\citep{Liu2023GroundingDINO}.
The backend was deployed on a \texttt{g4dn.xlarge} instance.

\begin{figure*}[t]
    \centering
    \includegraphics[width=\linewidth]{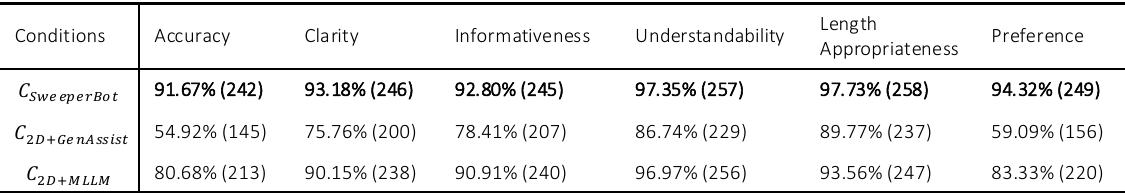}
    \caption{Percentage of descriptions created by \sysname~and baseline conditions positively rated by sighted participants in Study 2. The exact numbers of the descriptions are provided. }
    \label{fig::study2quant}
\end{figure*}

\section{Quantitative Summary of the Survey Response in Study 2}\label{sec::app::implementations}
This section presents additional details of our quantitative evaluations in Study 2 by sighted participants. 
Although Figure~\ref{fig::study2_rating} presents survey responses evaluating the quality of descriptions generated by \sysname~and two baseline conditions, Figure~\ref{fig::study2quant} shows how the percentage of descriptions positively\footnote{We consider ``excellent'', ``very good'', and ``good'' as positive ratings.} rated by sighted participants across six criteria (see Figure~\ref{fig::study2_rating}).
All quantitative results presented in Figure~\ref{fig::study2quant} can be computed from Figure~\ref{fig::study2_rating}.
Details of Study 2 can be referred to Section~\ref{sec::study::accuracy}.

%TC:endignore

\end{document}